\documentclass[twocolumn, floatfix]{aastex631}

\usepackage{newtxtext,newtxmath}
\usepackage{ulem}
\usepackage{gensymb}
\usepackage{array}
\usepackage{booktabs} 
\usepackage{graphicx}	
\usepackage{amsmath}	
\usepackage{textcomp}
\usepackage{breqn}
\usepackage[flushmargin]{footmisc}
\usepackage{xspace}

\shorttitle{}
\shortauthors{Vicentin et al.}

\graphicspath{{./}{figures/}}

\newcommand{\angstrom}{\textup{A\hspace{-1.2ex}\raisebox{1.2ex}{\scriptsize\ensuremath{\circ}}}\xspace}

\begin{document}

\title{Selecting Clusters and Protoclusters via Stellar Mass Density: I. Method and tests on Mock HSC-SSP catalogs. 
}

\correspondingauthor{Vicentin, Marcelo C.}
\email{marcelo.vicentin@usp.br}

\author[0000-0002-9191-5972]{Marcelo C. Vicentin}
\affil{Universidade de S\~ao Paulo, Instituto de Astronomia, Geof\'isica e Ci\^encias Atmosf\'ericas, Departamento de Astronomia, \\ SP 05508-090, S\~ao Paulo, Brasil}
\affil{Department of Astrophysical Sciences, Princeton University, Peyton Hall, Princeton, NJ 08544, USA}

\author[0000-0003-2860-5717]{Pablo Araya-Araya}
\affil{Universidade de S\~ao Paulo, Instituto de Astronomia, Geof\'isica e Ci\^encias Atmosf\'ericas, Departamento de Astronomia, \\ SP 05508-090, S\~ao Paulo, Brasil}

\author[0000-0002-3876-268X]{Laerte Sodré Jr.}
\affil{Universidade de S\~ao Paulo, Instituto de Astronomia, Geof\'isica e Ci\^encias Atmosf\'ericas, Departamento de Astronomia, \\ SP 05508-090, S\~ao Paulo, Brasil}

\author[0000-0002-0106-7755]{Michael A. Strauss}
\affil{Department of Astrophysical Sciences, Princeton University, Peyton Hall, Princeton, NJ 08544, USA}

\begin{abstract}

We present an algorithm designed to identify galaxy (proto)clusters in wide-area photometric surveys by first selecting their dominant galaxy—i.e., the Brightest Cluster Galaxy (BCG) or protoBCG—through the local stellar mass density traced by massive galaxies. We focus on its application to the Hyper Suprime-Cam Subaru Strategic Program (HSC-SSP) Wide Survey to detect candidates up to $\rm z \sim 2$. In this work, we apply the method to mock galaxy catalogs that replicate the observational constraints of the HSC-SSP Wide Survey. We derive functions that describe the probability of a massive galaxy being the dominant galaxy in a structure as a function of its stellar mass density contrast within a given redshift interval. We show that galaxies with probabilities greater than 50\% yield a sample of BCGs/protoBCGs with $\gtrsim 65\%$ purity, where most of the contamination arises from galaxies in massive groups below our cluster threshold. Using the same threshold, the resulting (proto)cluster sample achieves 80\% purity and 50\% completeness for halos with $M_{\rm{halo}} \geq 10^{14} \ M_{\odot}$, reaching nearly 100\% completeness for $M_{\rm{halo}} \geq 10^{14.5} \ M_{\odot}$. We also assign probabilistic membership to surrounding galaxies based on stellar mass and distance to the dominant galaxy, from which we define the cluster richness as the number of galaxies more likely to be true members than contaminants. This allows us to derive a halo mass–richness relation. In a companion paper, we apply the algorithm to the HSC-SSP data and compare our catalog with others based on different cluster-finding techniques and X-ray detections.

\end{abstract}

\section{Introduction} \label{sec:intro}

Galaxy clusters trace the densest regions of the universe on cosmic scales, and were formed by the collapse of perturbations of the primordial density field in the high-redshift Universe. These early dark matter halos grow by capturing other halos and baryonic matter, culminating in what we observe today as the most massive virialized structures. These clusters are observed at the ``nodes" of the cosmic web \citep[e.g.,][]{peebles80, bond96, crain15}.

The distinct characteristics of galaxy clusters make them compelling targets for testing models of large-scale structure growth and understanding the mechanisms driving galaxy evolution \citep{kravtsov12}. In the low redshift Universe ($\rm z \lesssim 1$), most galaxy clusters have virialized, establishing key properties such as concentrated distributions of galaxies, with the central region dominated by passive galaxies, and a predominantly hot, gaseous intra-cluster medium (ICM) detected through extended X-ray emission. These properties offer diverse avenues for cluster detection, primarily through identifying concentrations of red galaxies in optical and infrared observations \citep[e.g.,][]{bower99, chapman00, nakata01, oguri14, rykoff14, oguri18} and by detecting extended X-ray emission \citep[e.g.,][]{sarazin86, piffaretti11, koulouridis21, klein23, ota23}, which, in turn, enable the detection of the Sunyaev-Zeldovich effect \citep[e.g.,][]{sunyaev70, stanis09, bleem15, ade16, gobat19, hilton21, kitayama23}.

While the detection of extended X-ray emission and the Sunyaev-Zeldovich effect confirms the presence of a galaxy cluster, the homogeneous and efficient identification of cluster candidates across large volumes of the universe became possible through the advent of large multi-band photometric surveys in the optical and infrared. These surveys, covering large areas of the sky and with significant depth, offer an effective way to detect galaxy cluster and protocluster candidates homogeneously \citep[e.g.,][]{hao10, wylezalek13, rykoff16, gonzalez19, aguena21, qing22, werner23, doubrawa24}. Initiatives such as the Sloan Digital Sky Survey \citep[SDSS;][]{york00}, the Southern Photometric Local Universe Survey \citep[S-PLUS, ][]{mendesdeoliveira19}, the Hyper Suprime-Cam Subaru Strategic Program \citep[HSC-SSP, ][]{aihara18}, and the Dark Energy Survey \citep[DES, ][]{des05}, have been a valuable source of data for the identification and characterization of galaxy structures. Some algorithms, such as redMaPPer \citep{rykoff14} and CAMIRA \citep{oguri14}, focus on detecting the ``red sequence" observed in color-magnitude diagrams of galaxy clusters, primarily populated by the passive galaxies within these structures. Other methods rely solely on the distributions of galaxies in angular separation and photometric redshift \citep[e.g.,][]{wh12, wh15, aguena21}. As a general rule, these algorithms achieve a success rate of over 60\% at redshifts less than 1, when testing their algorithms with mock data.

However, a challenge lies in identifying (proto)clusters at higher redshifts. In the range $\rm 1 < z < 2$, a significant fraction of these structures is still in process of formation and are consequently referred to as protoclusters \citep[for a comprehensive review, see][]{overzier16}. These structures, not yet relaxed, are the progenitors of galaxy clusters that will eventually virialize, with masses greater than $\rm 10^{14} \ M_{\odot}$ at z = 0. Galaxies that, in a relaxed cluster in the local universe, would be within $\rm \lesssim 1 \ cMpc$, can be found dispersed over a region spanning several comoving megaparsecs at the protocluster stage \citep{chiang13}. Moreover, at these redshifts, the sample of galaxies with spectroscopy is considerably smaller and non-uniform, posing challenges to the estimation of photometric redshifts and/or color calibrations to determine red sequence galaxies \citep[e.g., ][]{oguri14}. Nevertheless, galaxy protoclusters are regions that already exhibit clumps with higher galaxy density ($\rm \gtrsim 4$$\sigma$) than the field \citep[e.g., ][]{harikane18, toshikawa18}, and galaxies in protoclusters show properties that already differ from field galaxies, suggesting that these galaxies undergo `pre-processing' before reaching the cluster stage \citep[e.g.,][]{chiang17, shimakawa18, werner22}.

In this paper, we introduce a novel algorithm for detecting galaxy cluster and protocluster candidates from optical imaging data, employing the identification of the dominant galaxy\footnote{Brightest Cluster Galaxy (BCG) for galaxy clusters and protoBCGs for galaxy protoclusters.} as the initial step. This approach differs from other methods already applied in the HSC-SSP footprint \citep{oguri18, wh21}, which typically detect the overall structure before identifying the dominant galaxy, offering a distinctive perspective on cluster identification \citep[e.g.,][]{eisenstein02}. BCGs are among the first galaxies formed in the dark matter halos of clusters and, as such, they possess unique characteristics when compared to other cluster or field galaxies \citep[e.g.,][]{bernardi09, lauer14}. BCGs are massive galaxies that already assembled half of their final stellar mass at z $\rm \gtrsim$ 1 mainly through multiple mergers with other gas-rich galaxies of similar mass (major wet mergers) and \textit{in situ} star formation, while at $\rm z \lesssim 1$, they have evolved primarily \textit{ex situ} by cannibalizing smaller, gas-poor galaxies (dry minor mergers), a process known as an \textit{inside-out} growth scenario \citep[e.g.,][]{delucia07, vandokkum10, montenegro23}. BCGs are giant elliptical galaxies with extended envelopes forming their halo and the intracluster light \citep[ICL, ][]{montes18, contini21}; they are the most massive galaxy within the cluster sitting near the bottom of the potential well; they are, in general, narrowly distributed in the high end of the cluster galaxy luminosity function and they tend to be uniform in color and redder than most other satellite galaxies \citep{postman95, lauer14, dalal21}. These distinctive features facilitate their identification among other galaxies in photometric surveys \citep[e.g.,][]{koester07}.

Our primary objective is to apply our algorithm to the HSC-SSP Wide Survey public data release 3 \citep{aihara22} to detect galaxy clusters and protoclusters up to $\rm z \sim 2$. The study of the HSC-SSP Wide Survey is justified by several characteristics, including its extensive coverage ($\rm \gtrsim 1000 \ deg^{2}$), its depth ($\rm i \sim 26$), and for having extensive overlap with other well-studied fields. The HSC-SSP wide provides data in five broad bands ($grizy$), limiting detections up to $\rm z \sim 1.4$, as beyond this redshift, the spectral break at 4000~\angstrom \xspace falls outside the wavelength coverage range. However, we aim to extend our analysis to higher redshifts by adding mid-infrared data W1 and W2 at 3.6 $\mu m$ and 4.5 $\mu m$, respectively, from the unWISE catalog \citep{schlafly18} when available, and high-accuracy photometric redshifts from the COSMOS2020 catalog \citep{weaver22} to estimate our own photometric redshifts, similar to the approach taken by \citet{wh21}. This redshift range is particularly intriguing, as it is the epoch when many protoclusters virialized to form clusters. Additionally, the identification of candidates in the redshift range $\rm 0.7 < z < 2$ holds particular significance due to the imminent commencement of operations of the Subaru Prime Focus Spectrograph \citep[PFS; for an overview, see][]{takada14, tamura16}. The PFS is poised to revolutionize our understanding of cluster formation by identifying and characterizing these structures at high redshifts. This instrument will be capable of measuring nearly 2400 spectra simultaneously in a single exposure, covering a hexagonal field of view with a diameter of 1.38 degrees. Furthermore, the PFS has extensive spectral coverage, spanning from the near-ultraviolet to the near-infrared (0.38 -- 1.26 $\mu m$).

We structure this paper as follows: in Section \ref{sec: algorithm}, we outline the functioning of our cluster finder algorithm. Next, in Section \ref{sec: data}, we introduce the simulation data we use to test and refine the algorithm, including details about the mock and other galaxy cluster candidate catalogs from the literature, which we utilize to assess the mock's consistency. Section \ref{sec: mock_prepare} presents the definitions adopted in the mock for detecting galaxy cluster and protocluster candidates, the procedures adopted for estimating photometric stellar mass and redshifts for PCcones objects, and consistency tests comparing mock data with the literature. Finally, in Section \ref{sec: application}, we elucidate our methodology for constructing probability models that determine the likelihood of a given galaxy being the dominant galaxy within a structure. Subsequently, we conduct a comprehensive assessment of our outcomes, considering both completeness and purity metrics derived from these probabilities. Furthermore, we define member selection criteria and determine Halo Mass-Richness relations. A summary is provided in Section \ref{sec: summary}. The results obtained from applying this algorithm to the HSC-SSP Wide Survey data, along with the photometric redshift estimation and other pertinent analyses such as comparisons with galaxy clusters identified by alternative algorithms, will be the subject of a forthcoming paper. Throughout this work we adopt a $\rm \Lambda$CDM concordance cosmology with $\rm h =$ 0.673, $\rm \Omega_{\rm m}$ = 0.315 and $\rm \Omega_{\rm \Lambda}$ = 0.685 \citep{planck1}.


\section{(Proto)Cluster finder algorithm}
\label{sec: algorithm}

The galaxy (proto)cluster finder algorithm presented here is designed to select, as a first step, dominant galaxies, i.e., BCGs or protoBCGs, within galaxy structures. Dominant galaxies are expected to be located at the core of their structures, where the mass density is higher than the outer regions of the structure itself and the field. This algorithm calculates the local stellar mass density associated with pre-selected massive galaxies (as described in Section \ref{sec: presel}) to select BCG or protoBCG candidates within a given galaxy structure through the stellar mass density contrast distribution.

For each pre-selected massive dominant galaxy candidate ($i$), we calculate the stellar mass located within a cylindrical volume centered on the candidate galaxy. This cylindrical volume is defined by a radius of $\rm r = 1 \ Mpc$ and a height corresponding to the comoving distance encompassed within the redshift slice 

\begin{equation}
    \Delta z_{i} = [z_i - \sigma_{z}(1 + z_i), z_i + \sigma_{z}(1 + z_i)],
    \label{eq: delta_z}
\end{equation}

\noindent
which depends on the redshift of the candidate accuracy, denoted by $\sigma_{z}$ (for our mock photometric redshift estimates, see Section \ref{sec: photoz}).

The density is measured by dividing this cylindrical volume into three equally spaced concentric rings. We apply a weighting factor ($w$) based on the inverse of the projected radial distance, i.e., $r^{-w}$, since it reduces the contribution of contaminants in the calculation. This means that galaxies closer to the dominant galaxy candidate have a higher weight in the density calculation. We chose $w = 0.8$, based on the completeness and purity analysis (see Section \ref{sec: effec} and \ref{sec: effec_cl}), to optimize our results. Finally, we averaged the densities obtained from all three rings to obtain an weighted average estimator,

\begin{equation}
    \hat{\rho}_{i} = \frac{\sum_{j = 1}^{3} \frac{M^{tot}_{\star i, j}}{\pi \ (r_{j}^{2} - r_{j-1}^{2}) \ d_{c}(\Delta z_{i})} \left( \frac{r_{j}}{cMpc} \right)^{-w}}{\sum_{j = 1}^{3} \left( \frac{r_{j}}{cMpc} \right)^{-w}},
    \label{eq: vol_cyl}
\end{equation}

\noindent
where $r_{j}$ and $r_{j-1}$ represent the outer and inner radii of the $j^{th}$ ring centered in the $i^{th}$ candidate, respectively, and $r_{0} \equiv 0$; $M^{tot}_{\star i, j}$ represents the total stellar mass contained within the ring; and $d_{c}(\Delta z_{i})$ is the radial comoving distance spanned within the redshift slice $\Delta z_{i}$.

Once we have the local stellar mass weighted average density associated with each massive galaxy, we calculate the density contrast as 

\begin{equation}
   \delta \rho_{i} = \frac{\hat{\rho}_{i} - \bar \rho}{\bar \rho},
   \label{eq: density_cont}
\end{equation}

\noindent
$\bar \rho$ is the reference density calculated as the average stellar mass density within the same $\Delta z_{i}$ and across the entire analyzed area (in this work, 36 deg$^2$ for each mock, see Section \ref{sec: pccones}).

As this algorithm is initially applied to lightcones emulating HSC-SSP wide observations, it is possible to construct a probabilistic model based on the distribution of density contrast for true mock dominant galaxies versus other pre-selected massive galaxies (see Section \ref{sec: presel}). Higher density contrast values are found to correlate with a higher probability of selecting dominant galaxies. Using this model, a threshold in density contrast can be defined, above which galaxies are more likely to be dominant. The model obtained through the simulated mock data then can be used to identify dominant galaxy candidates within observed data. We chose thresholds based on the expected number of galaxy clusters for a given redshift interval.

We identify (proto)cluster member galaxies by analyzing the photometric stellar mass and the distance to the dominant galaxy as functions of the photometric redshift, distinguishing between true (proto)cluster members and contaminants from background and foreground interlopers. Based on this information, for each selected dominant galaxy, we compute the probability of nearby galaxies, within a 1 Mpc radius, being true members or interlopers. Galaxies which have probabilities of being true members higher than a interloper according to both criteria are classified as members of the structure. The number of galaxies identified in this way defines the structure’s richness ($\lambda$). Since the mock datasets include information about the dark matter halo mass of the structures, we then establish halo mass-richness relations (see Section \ref{sec: cm_sel}).

In summary, this algorithm leverages local stellar mass density estimation and density contrast calculation to identify dominant galaxies and thus (proto)cluster candidates. Our approach does not rely directly on assumptions about the colors of galaxies in the (proto)cluster, and our first step is to identify the dominant galaxy rather than focusing on the galaxy (proto)cluster's structural features, which distinguishes our methodology from approaches employed by other cluster finders in the literature \citep[e.g., ][]{koester07, rykoff14, oguri14, wh21}. This methodological difference is particularly valuable at high redshifts, where protoclusters are more common. In these systems, the red sequence is often not yet well-defined. However, it is still possible to identify significant density contrasts associated with the core regions of these structures, where their dominant galaxies reside, enabling their detection \citep{overzier16, toshikawa18}.

\section{Data}
\label{sec: data}

In this section, we provide an overview of the data utilized in this project. This initial paper focuses on presenting the methodology of our galaxy cluster finder algorithm and its application to mocks that emulate observations from the HSC-SSP Wide Survey. First, we describe the mock lightcones \citep{araya24}--an updated version of \citep{araya21}--which serve as the primary dataset for optimizing selection criteria, probability modeling, and assessing the algorithm's efficiency in identifying dominant galaxies. Subsequently, we present a brief overview of catalogs of galaxy cluster candidates, obtained by other cluster finder algorithms, which will serve as an ancillary database for consistency checks of our mocks (see Section \ref{sec: cons_ch}).

\subsection{PCcones}
\label{sec: pccones}

The \texttt{PCcones} mock lightcones used in this project are generated using the \citet{henriques15} version of the \texttt{\texttt{L-GALAXIES}} semi-analytic model (SAM) applied to the Dark Matter-only Millennium simulation \citep{springel05} to generate galaxies. This SAM is designed to be run on the Millennium simulation merger trees obtained with the \texttt{SUBFIND} algorithm \citep{springel01}. This ensures that the galaxy formation and evolution processes simulated by \texttt{L-GALAXIES} are built upon merger trees containing information about the underlying dark matter distribution, resulting in an accurate representation of galaxy properties. The \texttt{\texttt{L-GALAXIES}} SAM is scaled to match the \citet{planck1} cosmological parameters using the \citet{angulo10} algorithm and incorporates a wide range of critical galaxy evolution processes, including gas infall and cooling, star formation, metal enrichment, supermassive black hole growth, and supernova and AGN feedback (for more details, refer to the Supplementary Material in \citealp{henriques15}). The SAM output provides physical properties for the synthetic galaxies, such as stellar mass, gas mass, and star formation rate.

The \texttt{Millennium} simulation has specific attributes that make it a good choice for this study. It features a dark matter particle mass of approximately $\rm 9.6 \/\ \times \/\ 10^{8} \/\ M_{\odot}/h$, allowing for the modeling of galaxies with stellar masses $\rm M_{\star} > 10^{8} \/\ M_{\odot}/h$. This range of stellar masses is particularly valuable for capturing a comprehensive view of galaxy populations, including those with lower stellar masses. Additionally, the simulation box size with $\rm L = 480.279 \/\ cMpc/h$ provides a volume large enough to accommodate a substantial number of galaxy structures. This is essential for our study, as it ensures that the simulation volume encompasses a diverse and representative sample of these cosmological structures, including the information of which galaxies belong to protoclusters. 

In this paper, we use the term \textit{structure(s)} for galaxies belonging to the same Friends-of-Friends (FOF) group in the Millennium simulation with a mass exceeding $\rm 10^{14} \ M_{\odot}$ at the desired redshift or those that will surpass this mass threshold in their future evolution, representing galaxy clusters or protoclusters, respectively. FOF groups are identified in the Millennium simulation as groups of dark matter particles that lie within one-fifth of the mean inter-particle distance from each other \citep{davis85}. The mass of the FOF group is quantified using the \texttt{m_tophat} parameter provided by the simulation, which represents the total dark matter mass enclosed within a radius where the overdensity corresponds to the virialization threshold in the top-hat collapse model, consistent with the cosmology adopted in this work \citep{planck1}. Here, we will use the term $M_{\rm{halo}}$ instead of \texttt{m_tophat} to refer to the dark matter halo mass of the structures.


\texttt{PCcones} lightcones estimate spectro-photometric properties of synthetic galaxies following the post-processing approach outlined by \citet{shamshiri15}. This was made using \texttt{L-GALAXIES} star formation history (SFH) output for each galaxy. Leveraging this SFH data, they attribute spectral energy distributions (SEDs) to individual stellar populations within each SFH bin. For consistency with the \texttt{Millennium} lightcones, we employed SED templates from the \citet{maraston05} stellar synthesis population models, assuming a \citet{chabrier03} initial mass function. The dust extinction was included following \citet{henriques15}. The \texttt{PCcones} magnitudes were computed from redshifted SEDs based on the \citet{shamshiri15} post-processing description. The SEDs for each galaxy also allow us to compute the apparent magnitude emulating the filter response of the desired instrument. For the purposes of this study, we will make use of the optical $grizy$ Subaru's HSC \citep{kawanomoto18} and IRAC 3.6 and 4.5 $\mu m$, including K-correction. As shown in \citet{araya21}, the post-processing approach to obtain observer-frame magnitudes presents reliable results when computing photometric redshifts using mock data with conventional photo-z algorithms.

Once the merger tree information of the dark matter halos from the Millennium simulation and their corresponding masses are available, the PCcones mock provides the information of which structures have already reached or are expected to surpass the cluster halo mass threshold of $M_{\rm{halo}} = 1 \times 10^{14} \ M_{\odot}$ at some future point in the simulation for $\rm z > 0$. This enables the classification of galaxies as inhabitants of clusters or protoclusters.

To have a significant volume to identify galaxy clusters robustly, our analysis in this project makes use of 10 lightcones with a fixed area of 36 $\rm deg^{2}$ each, in which we selected galaxies with $\rm i < 25.5$ emulating the HSC-SSP Wide completeness limit. We verified that key statistics such as the cluster number density show stable behavior as the number of lightcones increases, demonstrating that 10 mocks is sufficient for the goals of this work.

In order to create a mock dataset including observational realities, we introduced errors in magnitudes. The magnitude errors were derived by fitting the HSC-SSP Wide Survey errors \citep{bosch18} as a function of the respective magnitudes in different bands\footnote{We are using \texttt{cmodel} magnitudes and their respective errors.} and also for W1 and W2 from unWISE. We selected only objects with \texttt{\textit{filter}\_extendedness\_value} = 1 in all five HSC filters, which were classified as extended sources in the HSC-SSP database. The fits were obtained using an exponential function (Eq. \ref{eq: magerr_func})

\begin{equation}
    e_{m}^{2}(m; \gamma_{1}, \gamma_{2}, m_{5\sigma}) = \gamma_{1} \times 10^{\gamma_{2} \/\ \times \/\ (m - m_{5\sigma})}, 
    \label{eq: magerr_func}
\end{equation}

\noindent
where $m$ denotes the magnitude in a given band; $\gamma_{1}$ and $\gamma_{2}$ are fitted parameters; and $m_{5\sigma}$ is the 5-$\sigma$ limit magnitude in the same band. We use a Markov Chain Monte Carlo (MCMC) method to determine the optimal parameters. Table \ref{tab: error_pars} presents the parameters obtained for each band.

Based on these fits, the mock magnitudes ($m_{true}$) were perturbed as 

\begin{equation}
    m_{pert} \sim \mathcal{N}(\mu = m_{true}, \/\ \sigma = e_{m})
    \label{eq: pert_mag}
\end{equation}

\begin{table}
\centering
\caption{Error function parameters obtained for each filter.}
\begin{tabular}{cccc}
\hline
Filter & $m_{5\sigma}$ & $\gamma_1$ & $\gamma_2$ \\
\hline
g & 27.510 & 0.451 & 0.761 \\
r & 26.988 & 0.380 & 0.798 \\
i & 25.499 & 0.029 & 0.993 \\
z & 26.172 & 0.461 & 0.843 \\
y & 25.635 & 0.576 & 0.709 \\
W1 & 22.991 & 0.146 & 0.757 \\
W2 & 22.182 & 0.166 & 0.726 \\
\hline
\end{tabular}
\label{tab: error_pars}
\end{table}

\subsection{CAMIRA HSC-SSP wide galaxy cluster candidates}

Using the Cluster finding algorithm based on Multi-band Identification of Red sequence gAlaxies (CAMIRA) cluster finder algorithm \citep{oguri14}, \citet{oguri18} identified galaxy cluster candidates in $\rm \sim 232 \/\ deg^{2}$ of the HSC-SSP wide photometric survey in the internal HSC survey data release ``S21A'' up to redshift 1.1. In this study, we will use various physical properties of galaxies identified by CAMIRA as BCGs to compare them with galaxies defined as BCGs in our mocks (see Section \ref{sec: cons_ch}). In broad terms, the CAMIRA algorithm identifies concentrations of red galaxies with colors compatible with the red sequence by modeling the SEDs of galaxies using the calibrated SPS model of \cite{bruzual03}. For each galaxy, the algorithm calculates the likelihood of it belonging to the red sequence at a given redshift. These values are then employed to generate richness maps by applying stellar mass and spatial filters in specific redshift slices. Peaks identified in these maps are considered galaxy cluster candidates. The algorithm leverages the relative positions of these peaks and galaxies to assess the likelihood of galaxies belonging to the structure. It estimates the redshift of the structure by analyzing the photometric redshift estimates of galaxies and their positions relative to the peak. Potential BCGs are identified by evaluating the stellar masses of galaxies and their distance from the peak. The position of the BCG candidate is adopted as the center of the structure, and the redshift is recalculated. This process continues iteratively until convergence.

\subsection{Wen \& Han 2021 HSC-SSP wide galaxy cluster candidates}

\citet{wh21} (W\&H21) present an application to $\rm \sim 800 \/\ deg^{2}$ of HSC-SSP Wide imaging data of their cluster finder algorithm \citep{wh12,wh15}. Only galaxies with cross-matches in the mid-infrared unWISE survey were used, totaling 14.68 million galaxies. With this setup, they performed their own estimation of photometric redshifts and identified galaxy cluster candidates in the redshift range $\rm 0.1 < z < 2$. The algorithm detects galaxy cluster candidates by grouping luminous galaxies within a 0.5 Mpc projected radius and in the same photometric redshift slice, using a FOF approach \citep{huchra82}. For each galaxy in the sample, they count the number of luminous galaxies within 0.5 Mpc projected radius. The temporary center of each group is set as the position of the galaxy with the largest number of friends. From this center, galaxies within a 1 Mpc projected radius and in the same redshift slice are considered group members, with the redshift of this structure estimated as the median of their photometric redshifts. Subsequently, the algorithm identifies the brightest cluster galaxy (BCG) as the brightest galaxy within a 0.5 Mpc radius of the temporary center, recalculating cluster properties with the BCG as the center. A galaxy cluster candidate is selected if the total luminosity of the FOF exceeds a predefined threshold. Finally, potential duplicate clusters are merged using the FOF algorithm, considering photometric redshift and projected separation.

\subsection{redMaPPer galaxy cluster candidates}

The redMaPPer algorithm \citep{rykoff14} is designed for deep wide-field photometric cosmology surveys, aiming to identify overdensities of galaxies based on the probabilities of these galaxies belonging to the red sequence at a given redshift. They assign them as central or satellite members with a probabilistic approach. Validated against X-ray and Sunyaev–Zel’dovich observations, the algorithm counts the excess number of red-sequence galaxies within a specific radius and brightness threshold to determine cluster richness. Each cluster is centered on the most likely central galaxy based on brightness, richness, and local density. Additionally, each red-sequence cluster member is assigned a membership probability. In \citet{rykoff14}, the redMaPPer algorithm was applied to the BOSS region \citep{dawson13} covering 10,400 deg² from the SDSS DR8 photometric catalog \citep{aihara11}. By utilizing red galaxy spectroscopic redshifts, a robust red sequence model was established to define both richness and photometric redshift estimators. Photometric redshifts exhibit small bias and low scatter, ranging from $\sigma_{z}$ = 0.006 to 0.020 from z $\rm \sim$ 0.1 to 0.5, respectively. The richness threshold was set at 20 detected red sequence galaxies to ensure robustness, with an associated halo mass threshold of $\rm M \geq 10^{14} \ M_{\odot}$. The performance of the algorithm is expected to decline at lower richness levels.

\section{Preparing mock data for cluster finding}
\label{sec: mock_prepare}

In this section, we will discuss the data preparation process for the PCcones mock dataset. Our approach involves the identification of dominant galaxies and leveraging the information about galaxy clusters and protoclusters, as their halo masses, the structure's member galaxies and their stellar masses. We will start by outlining the definitions employed to categorize different types of galaxies within the mock dataset (Section \ref{sec: definitions}). To ensure that the simulated data closely resembles real observations, we incorporate photometric redshift and stellar mass estimates into the mock dataset (Section \ref{sec: photoz}). These estimated quantities are used exclusively in all subsequent analyses throughout this paper. Furthermore, in Section \ref{sec: cons_ch}, we will utilize observational catalogs to assess the consistency of our galaxy classifications.

\subsection{Definitions}
\label{sec: definitions}

As part of our definitions, we will establish criteria to determine which galactic structures constitute a galaxy cluster or a protocluster. To achieve this, we will rely on the total mass of halos (dark matter-only) within the same FOF group from the Millennium simulation, adopting a halo mass threshold of $M_{\rm{halo}} = 10^{14} \ M_{\odot}$ at the redshift in consideration. In other words, galaxies that belong to halos within the same group, where the sum of their masses exceeds this threshold, will be categorized as \textit{cluster members}. If the total halo mass is below this threshold but will surpass it before z = 0, the structure will be classified as a galaxy protocluster. To streamline our subsequent analysis, we opt not to retain information about galaxy groups—structures that have mass below $\rm 10^{14} \ M_{\odot}$ at the desired redshift and fail to meet the mass threshold at any point in the simulation's future. Although this definition relies on information only available in simulations, our algorithm offers a potential observational proxy to distinguish such systems. As shown in Figure \ref{fig: probs}, dominant galaxies in protoclusters reside in significantly denser regions than those in groups. Consequently, as demonstrated in Figure \ref{fig: sel_objs}, our selection is efficient in identifying (proto)BCGs, while failing to recover the so-called Brightest Group Galaxies (BGGs), which, under our framework, are classified as field galaxies. 

While the primary focus of this study is the identification of galaxy clusters, we also aim to retain information about protoclusters since this enables us to quantify, among our identifications, which objects belong to this category of structures.

\begin{figure}

	\includegraphics[width=\columnwidth]{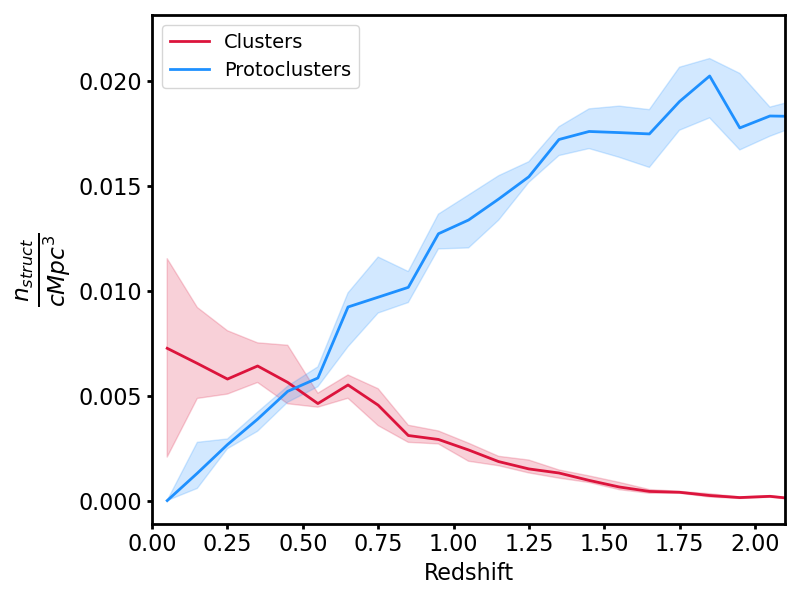}
    \caption{Number density of structures as a function of redshift in the PCcones mocks. Red (blue) line denotes the median number density galaxy clusters (protoclusters), while the shaded area is limited by 16th and 84th percentiles, considering 10 different lightcones with $\rm 36 \ deg^{2}$ each. We are defining galaxy clusters as structures with halo mass $M_{\rm{halo}} \geq 10^{14} \ M_{\odot}$ and galaxy protoclusters as structures with halo mass below this threshold, which will surpass this value at some point in their future at z $>$ 0. The large error bars at low redshifts for clusters are primarily due to cosmic variance. At low redshift, the survey volume per unit redshift is smaller, which enhances the impact of large-scale structure fluctuations on the measured cluster number densities.}
    \label{fig: n_cl_prot}
\end{figure}

Figure \ref{fig: n_cl_prot} depicts the number density of structures as a function of redshift. The volume calculations were made using steps of 0.1 in true mock redshift. We have also constrained the plot to cover the redshift range up to $\rm z = 2$, as this is the redshift interval within which we will search for galaxy structures. Notice that here we are using the true mock redshift.

The plot shows that the density of galaxy clusters gradually increases from $\rm 2 \times 10^{-4}$ to $\rm \sim 7 \times 10^{-3}$ clusters per cubic comoving megaparsec from $\rm z \sim 2$ to z = 0. This suggests that these structures predominantly start forming at $\rm z < 2$, with rare cases at higher redshifts. On the other hand, protoclusters exhibit a smooth decline between redshifts 2 and 1.3, and, subsequently, their density decreases more rapidly, virtually approaching zero at $\rm z \sim 0.05$, by definition. The rare cases of protoclusters at low redshifts ($\rm z < 0.2$) refer to massive galaxy groups with a total halo mass close to our mass limit for defining clusters of $\rm 10^{14} \ M_{\odot}$. These structures, at some point at z $>$ 0, reach this mass limit in the simulation and, therefore, are classified as protoclusters.

\begin{figure}

	\includegraphics[width=\columnwidth]{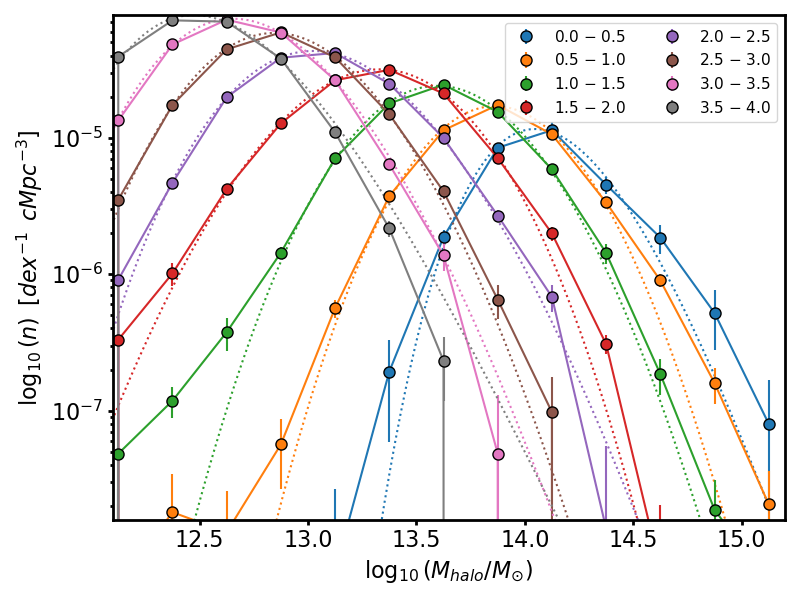}
 	\includegraphics[width=\columnwidth]{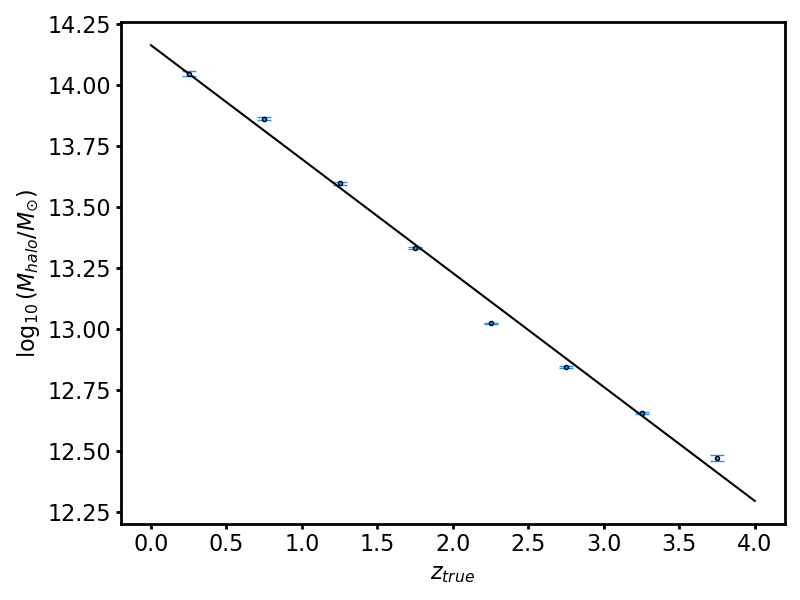}

    \caption{Top: Structure's halo mass functions for different redshift intervals depicted in the legend. The solid line is simply connecting the points, while the dotted lines are modified Normal distributions fits. Bottom: Halo mass as a function of the redshift. Each point denotes the halo mass where the modified Normal distributions reach their peak. The line denotes a power-law fit with 3 free parameters to these data. The function with the fitted parameters is showed in the upper right of this panel.  
}
    \label{fig: struct_mass_funcs}
\end{figure}

The top panel of Figure \ref{fig: struct_mass_funcs} illustrates the halo mass functions of galaxy clusters and protoclusters present in the mock for different redshift bins, as shown in the legend. The solid lines simply connect the points, while the dotted ones are Normal distribution fits with an extra parameter for the shape (s), which introduces skewness by allowing the width to vary with the distance from the mean, described as 

\begin{equation}
    f(x; A, \mu, \sigma, s) = A \exp\left\{ -\frac{(x - \mu)^2}{2[\sigma + \text{{s}}  (x - \mu)]^2} \right\},
\end{equation}

\noindent
where the input $x$ is defined as $x = \log_{10}(M_{\rm{halo}}/M_{\odot})$. 

The evolution of the mass function is clearly evident from these distributions. Although the modified Normal distributions do not fit the data well at points far from the means, around the peak, where $n \rm \gtrsim 10^{-6} dex^{-1} Mpc^{-3}$, the fits closely match the data. This allows us to examine how this peak evolves, as depicted in the bottom panel, where each point represents the halo mass at the peak of each distribution and the solid line is a simple power-law fit. In the highest considered redshift range ($\rm z = 3.5 - 4$), we observe the peak at $\log(M_{\rm{halo}}/M_{\odot}) = 12.5$, which evolves, at $\Delta \log(M_{\rm{halo}}/M_{\odot}) \sim 0.25$ intervals, to $\rm 14.125$ in the lowest redshift interval. Analyzing the evolution of the last point in each distribution in the upper panel, representing the most massive structures for each redshift bin, reveals few cluster-sized structures at $\rm z > 2$. The most massive structures with $\log(M_{\rm{halo}}/M_{\odot}) \geq 15$ only emerge at $\rm z < 1$, and even at lower redshifts, they remain exceptionally rare\footnote{Some observational works report higher mass estimates for protoclusters at $z > 2$, often based on overdensities and assumptions about future merging of multiple clumps into a single system by z = 0 \citep[e.g.,][]{cucciati18, toshikawa18, steidel00}.}.

Identifying the dominant galaxy within each structure (BCG or protoBCG) is pivotal for our approach to structure identification, as this is our first step to find galaxy (proto)clusters. Therefore, we define the BCG (protoBCG) of a cluster (protocluster) as the member galaxy with the highest stellar mass. All other galaxies which belong to (proto)clusters are defined as \textit{Cluster Member}. Galaxies not classified as \textit{BCG}, \textit{protoBCG}, or \textit{Cluster Member}, are classified as \textit{Field} galaxies.

In the next subsection, we will present the procedures adopted to estimate photometric redshifts and stellar masses of galaxies.

\subsection{Photometric redshifts and stellar masses}
\label{sec: photoz}

To estimate photometric stellar masses and redshifts for the galaxies in mocks, we used a fully-connected feed-forward neural network (see \citealp{vicentinII25} for a detailed description). It comprises an input layer with 15 neurons, encompassing the $grizy$ magnitudes along with their estimated errors simulating HSC-SSP wide data, and $W1$ and $W2$ magnitudes with their respective estimated errors simulating unWISE photometry (Section \ref{sec: pccones}). In the case of stellar mass estimation, the input layer has an extra neuron corresponding to the estimated photometric redshift, totaling 16 input features. Additionally, the network includes a linear output layer with one neuron (representing either the predicted redshift or the stellar mass) and four hidden layers activated by Rectified Linear Unit (ReLU) with 256 neurons each. The Keras package \citep{chollet18} was chosen as the library for this project. 

\begin{figure*}

	\includegraphics[width=\textwidth]{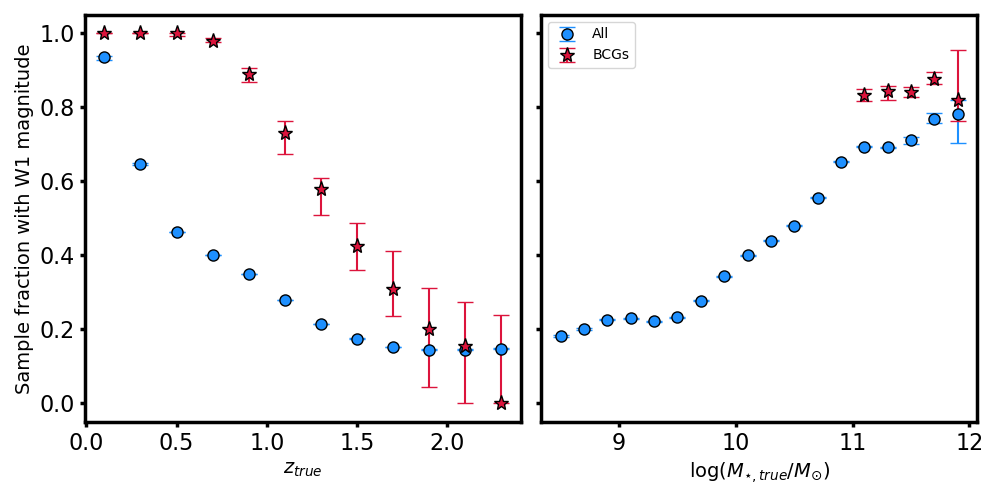}

    \caption{Sample fraction with measured W1 values as a function of the true redshift (left panel) and true stellar mass (right). Blue dots denote all mock objects while red stars represent BCGs.
    }

    \label{fig: W1_objs}
\end{figure*}

To emulate the observations more realistically, we removed measurements in the $W1$ and $W2$ bands randomly in i-band magnitude bins, to match the fraction of galaxies that do not have these measurements in the observations. Unlike W\&H21 who use only galaxies with complete photometry in $W1$, we kept all the objects in our sample including objects without photometry in the unWISE filters. Figure \ref{fig: W1_objs} shows the fraction of galaxies with W1 photometry as a function of true redshift (left panel) and true stellar mass (right panel). The fraction of objects with W1 measurements decreases rapidly with redshift, reaching $\rm \sim 18\%$ at $\rm z_{\text{true}} \gtrsim 1.5$. Conversely, nearly all BCGs have W1 values up to $\rm z_{\text{true}} \sim 0.7$; the fraction rapidly decrease to $\rm \sim 20\%$ at $\rm z_{\text{true}} \sim 2$. This behavior can be explained by examining the right panel, where more massive galaxies have a higher fraction with W1. Even for the most massive objects with $\rm \log(M_{\star}/M_{\odot}) > 11$, the group of BCGs has a higher fraction than when considering all objects in the mock within the same mass interval. The number density of BCGs at high redshifts ($\rm 1.5 \lesssim z_{\text{true}} < 2$) is much lower than at low redshifts (see Figure \ref{fig: n_cl_prot}).

\begin{figure*}

	\includegraphics[width=\textwidth]{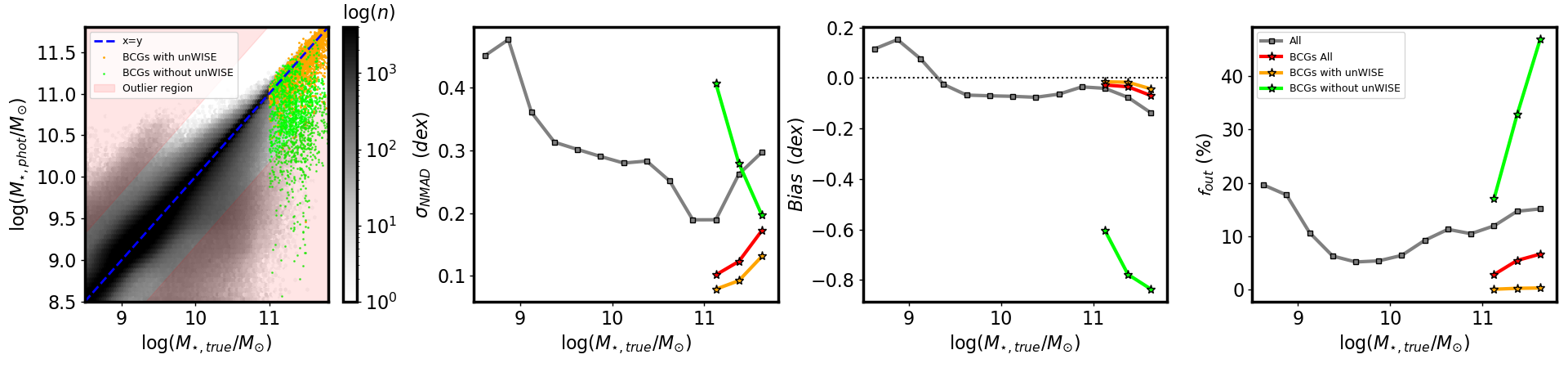}
	\includegraphics[width=\textwidth]{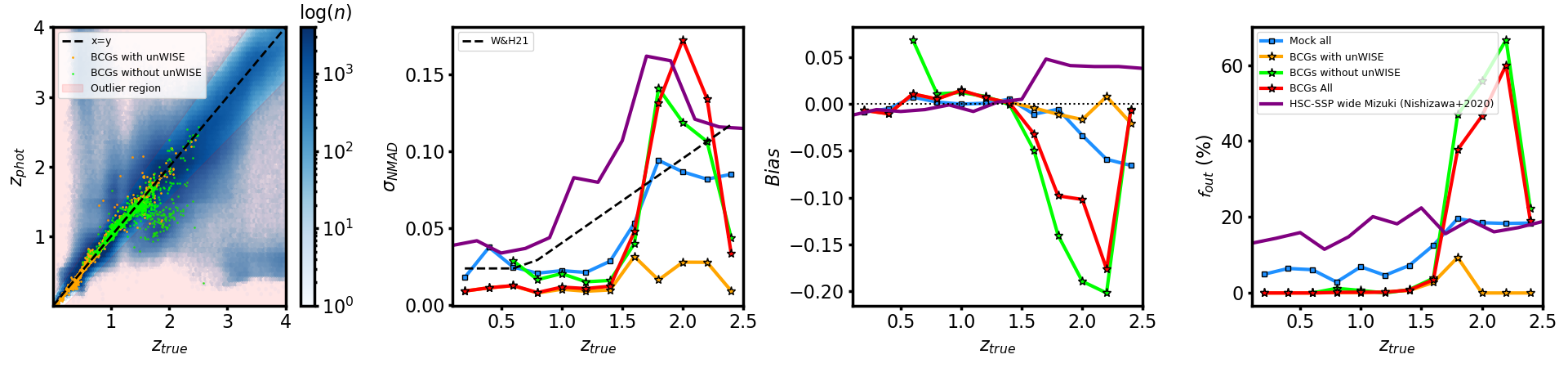}
	\includegraphics[width=\textwidth]{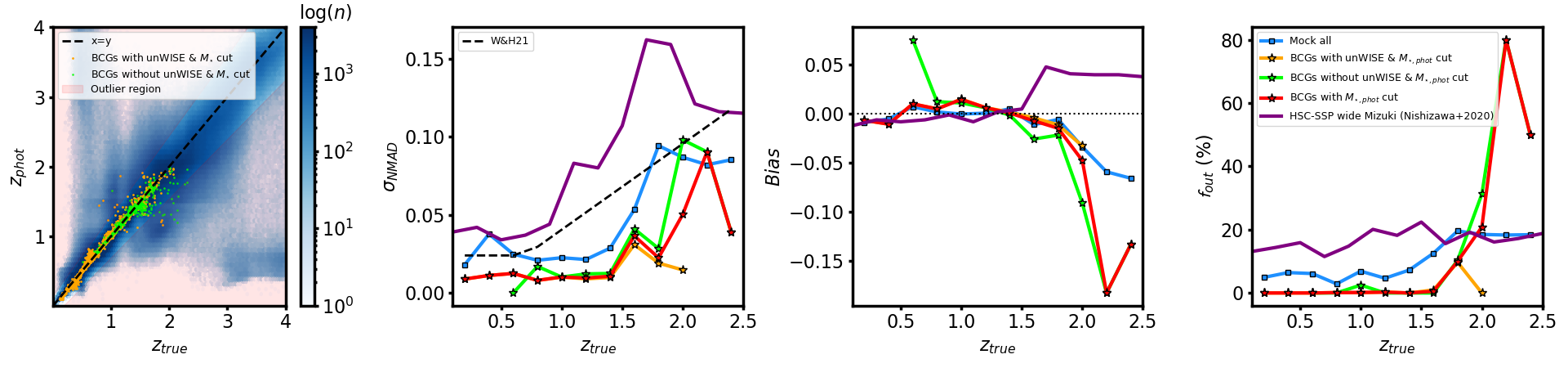}

    \caption{From left to right: the first row of plots shows the estimated stellar mass ($\rm M_{\star, phot}$) of galaxies, $\sigma_{NMAD}$ (Eq. \ref{eq: nmad}), $Bias$ (Eq. \ref{eq: bias}), and $f_{out}$ (Eq. \ref{eq: f_out}), respectively, as a function of the mock stellar mass ($\rm M_{\star, true}$). Gray dots and lines represent all objects from the test sample. Orange (green) dots and lines stand for true mock BCGs with (without) W1 unWISE band information. Red lines stand for all true mock BCGs. Second row of plots shows the results for photometric redshift estimates ($\rm z_{phot}$) as a function of $\rm z_{true}$, analogous to the stellar mass estimates in the first row. Blue dots and lines denote all objects in the test sample. The purple line represents the results obtained by \citet{nishizawa20}. The dashed line in the second plot ($\sigma_{NMAD} vs. z_{true}$) denotes the results obtained by W\&H21. Third row of plots is analogous to the second row, now including only BCGs after pre-selecting galaxies above a given threshold in photometric stellar mass (as described in Section \ref{sec: presel}).
}

    \label{fig: photoz}
\end{figure*}

We randomly selected $\rm 8 \times 10^{5}$ mock galaxies with $\rm i \lesssim 25.5$ (a similar number to what we have for training the HSC-SSP observational spectroscopic sample), regardless of whether they are field galaxies or cluster members, as our training (80\%) and testing (20\%) sample to estimate photometric stellar masses and redshifts. Figure \ref{fig: photoz} presents the results for our photometric stellar mass and redshift estimates for mock data. The left most plot shows the estimated quantity ($\rm \log(M_{\star, phot}/M_{\odot})$ or $\rm z_{phot}$) as a function of the mock true quantity, while the other three plots display three different metrics evaluating the quality of the estimates as a function of the true value of the quantity \citep[e.g.,][]{lima22}. The metrics measure dispersion using the normalized median absolute deviation \citep[$\sigma_{NMAD}$, Eq. \ref{eq: nmad}; ][]{hoaglin83}, systematic deviations ($\rm Bias$, Eq. \ref{eq: bias}), and the outlier fraction ($\rm f_{out}$, Eq. \ref{eq: f_out}):

\begin{equation}
    \sigma_{NMAD} (z_{true}) = 1.48 \times median \left(\frac{\delta z - median(\delta z)}{1 + z_{true}}\right),
    \label{eq: nmad}
\end{equation}

\noindent
where $z_{true}$ is the mock redshift, $z_{phot}$ is the estimated photometric redshift, and $\delta z = z_{phot} - z_{true}$. 

\begin{equation}
    Bias (z_{true}) = median \left(\frac{\delta_{z}}{1 + z_{true}}  \right);
    \label{eq: bias}
\end{equation}

\begin{equation}
    f_{out} (z_{true}) = \frac{N_{out}}{N_{tot}} 
    \label{eq: f_out},
\end{equation}

\noindent
where $N_{out}$ is defined as the number of objects that satisfy the condition $\frac{|\delta z|}{(1 + z_{true})} \geq 0.15$, and $N_{tot}$ is the total number of objects. Analogous metrics were adopted for evaluating the stellar mass estimates.
\\

The first row of plots in Figure \ref{fig: photoz} displays the results obtained for the photometric stellar mass estimates. Not surprisingly, BCGs with unWISE bands information (orange) have considerably better photometric stellar mass estimates than those without (green), note that the coverage range with the HSC bands no longer encompasses the 4000 \angstrom \xspace break at $\rm z \gtrsim 1.4$. When considering all BCGs (red lines), we observe that the metrics are much closer to the values obtained for the BCG sample with unWISE bands information. This can be explained by examining the right plot of Figure \ref{fig: W1_objs}, which illustrates that over 80\% of the BCGs have W1 detections across all photometric stellar mass intervals.

The second row of plots in Figure \ref{fig: photoz} displays the results obtained for the photometric redshift estimates. We also included the results obtained by \citet{nishizawa20} for the HSC-SSP wide using the template fitting method Mizuki \citep{tanaka18}. In the second plot ($\sigma_{NMAD} \/\ vs. \/\ z_{true}$), the dashed line represents the result obtained by \citet{wh21}, who only use objects with complete photometry in $grizyW1$. In the interval $\rm 1.5 < z_{\text{true}} < 2$, there is a degradation in the estimates overall, which is expected mainly due to the smaller number of objects with values in the unWISE bands (Figure \ref{fig: W1_objs}). Those BCGs without detections in W1 have considerably worse estimates in this redshift interval. The red curve, representing the metrics considering all BCGs, is much closer to the green curve representing the BCGs without measurements in unWISE. The left panel of Figure \ref{fig: W1_objs} helps us understand the reason behind this, as in this redshift interval, the fraction of BCGs with unWISE measurements is between 20 and 40\%. 

The third row of plots in Figure \ref{fig: photoz} is similar to the plots in the second row. However, in this case, we are considering only the BCGs selected above a given threshold in photometric stellar mass. In Section \ref{sec: presel}, we explain the procedure adopted to obtain these thresholds. For this case, BCGs removed from the sample are generally those with poorer quality estimates without unWISE detections. That is, this cut gives significantly better results.

\subsection{Consistency checks}
\label{sec: cons_ch}

\begin{figure*}

	\includegraphics[width=\textwidth]{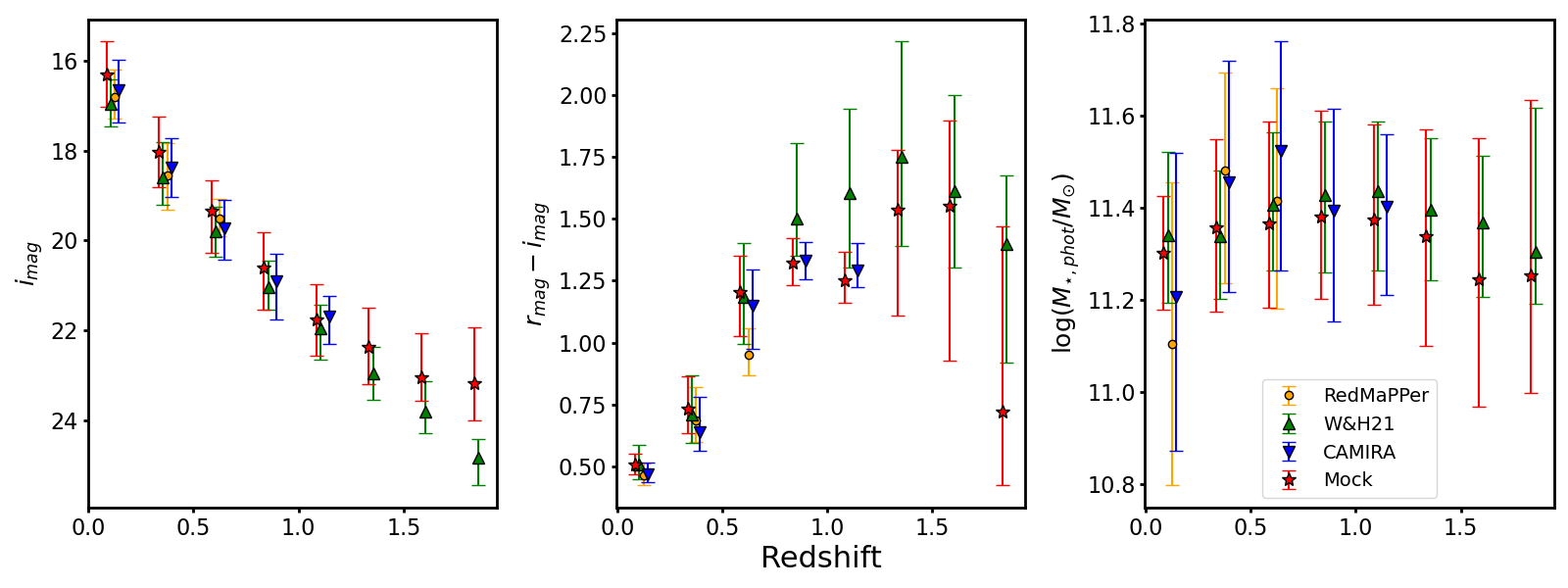}
    \caption{BCG $i$-band magnitude, $r - i$ observed frame color, and stellar mass as function of redshift. For the mocks, we utilized perturbed magnitudes and photometric stellar masses and redshifts (Sections \ref{sec: pccones} and \ref{sec: photoz}) for true mock BCGs. Points denote the median properties within a given redshift bin, and the bars are the dispersions bounded by the 16th and 84th percentiles. BCGs from CAMIRA \citep{oguri14}, W\&H21 \citep{wh21}, and redMaPPer \citep{rykoff14} catalogs are represented in the plots by blue, green, and orange colors, respectively, while mock BCGs are denoted by red color. Also, redMaPPer galaxies were cross-matched with the HSC-SSP Wide Survey to compare the measurements in the same photometric system. A slight horizontal shift was applied to the different markers to improve visualization.
}

    \label{fig: bcg_lit_comp}
\end{figure*}

In Section 3 of \citet{araya21}, several validation tests are conducted with PCcones mocks, comparing them with various observations, including data from the HSC-SSP survey. Statistical consistency is observed in galaxy number counts as a function of \textit{griz} magnitudes. The color-magnitude diagrams are also in good agreement with observations. For the purpose of this work, we will incorporate two additional validation tests: the distributions of BCG properties as a function of redshift (Figure \ref{fig: bcg_lit_comp}) and a comparison of the radial profiles of clusters and the velocity dispersion distribution with the SDSS redMaPPer cluster sample \citep{rykoff14}. To ensure a fair comparison with the latter, we included the same magnitude limit and selected mock galaxy clusters within the same redshift limits as the redMaPPer cluster sample.

Figure \ref{fig: bcg_lit_comp} presents the $i$-band magnitude, $r - i$ observed frame color, and stellar mass of the BCGs as function of redshift. We utilized perturbed magnitudes and photometric stellar masses and redshifts, as detailed in Sections \ref{sec: pccones} and \ref{sec: photoz}. For comparison, three observational samples of BCGs obtained using cluster finder algorithms in optical data (described in Section \ref{sec: data}) have been included: CAMIRA, W\&H21, and redMaPPer. We cross-matched redMaPPer galaxies with HSC-SSP Wide Survey to compare the measurements in the same photometric system.

There is consistency within the percentiles across the entire analyzed redshift range between true BCGs in the mock data and those from other catalogs. Only the W\&H21 catalog includes BCGs at redshifts above $\rm \sim$ 1.3. For BCGs above $\rm z \sim 0.7$, objects in the W\&H21 catalog tend to have redder colors than those in the mock or CAMIRA catalogs. At $\rm z > 1.2$, BCGs in the mock dataset exhibit slightly higher stellar masses than the W\&H21 catalog. Nevertheless, these differences are not statistically significant.

\begin{figure}

	\includegraphics[width=\columnwidth]{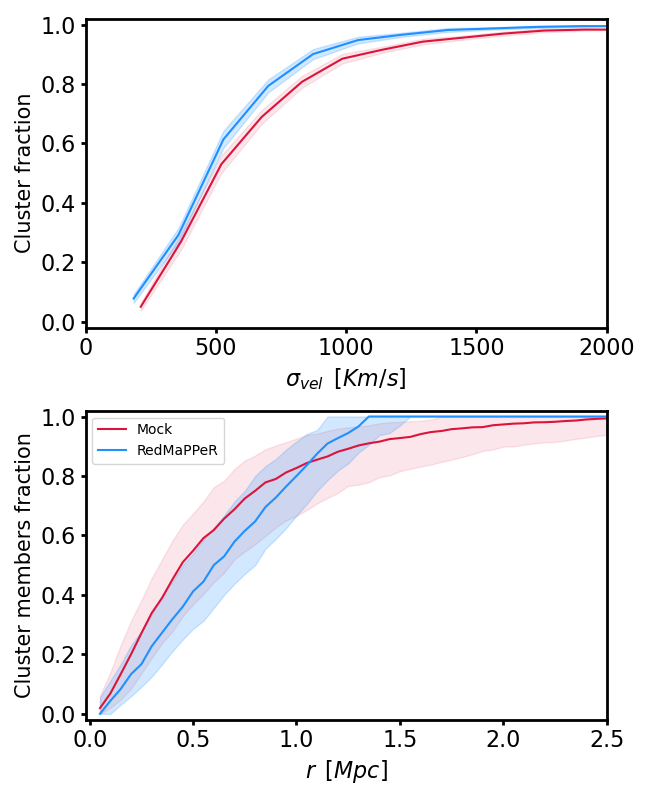}
    \caption{Red (blue) curve denotes mock (RedMaPPer) cumulative statistics for true mock galaxy clusters selected within the same magnitude limit ($\rm i_{mag} < 21$) and at the same observed redshift interval ($\rm z < 0.8$) as in the RedMaPPer sample from \citet{rykoff14}, used here for comparison. Upper plot: Distribution of true mock galaxy clusters as a function of the velocity dispersion, calculated following \citet{ferragamo20}. Bottom plot: Cumulative radial profile of the true mock clusters relative to their BCGs. Shaded area encompasses the 16th and 84th percentiles. In both cases we are using the BCG 3D coordinates (RA, Declination, and redshift) as the center and the distance to the other galaxies is calculated as the angular diameter transverse distance.
}

    \label{fig: struct_consist}
\end{figure}

Figure \ref{fig: struct_consist} presents two plots that assess the 3D spatial distribution of galaxy cluster member galaxies in the mock dataset (red curves) and the redMaPPer catalog (blue curves). Assuming that the BCGs occupy the center of the clusters, in the top plot, we depict the fraction of galaxy clusters as a function of the velocity dispersion ($\sigma_{vel}$), calculated using the bi-weight estimator as presented in \citet{ferragamo20}. This plot aims to evaluate the distribution of member galaxies in redshift space with respect to the dominant galaxy. In the bottom plot, we illustrate the fraction of member galaxies as a function of the distance from the cluster center, allowing us to assess the angular diameter transverse distance of member galaxies relative to the BCG.

To compare whether there is a statistical difference between the velocity dispersions of the mock dataset and the clusters from redMaPPer, we performed a two-sided Kolmogorov-Smirnov (KS) test. This test checks whether the underlying continuous distributions of the two datasets can be considered consistent with each other. The obtained p-value was 0.245, indicating that the samples are statistically similar.

In the bottom plot of Figure \ref{fig: struct_consist}, the shaded area encompasses the 16th and 84th percentiles of the radial profile of the clusters. For the redMaPPer galaxies we use spectroscopic redshifts when available. If not, we use their photometric redshifts (see Section \ref{sec: data}). There is consistency between the structures in both catalogs. The mock curve is smoother beyond 1 Mpc than the observed curve. This is likely related to the fact that, in the mock datasets, we know all the galaxies that are cluster members, whereas the observational samples are constrained by the method used to determine which galaxies are members. This determination is often associated with the central regions of the structure, where there is a higher density of member galaxies and, therefore, a lower contamination by interlopers.

\section{Application of the algorithm to mock data}
\label{sec: application}

In this section, we present the process of applying the algorithm described in Section \ref{sec: algorithm} to the PCcones mock lightcones described in Sections \ref{sec: pccones} and \ref{sec: mock_prepare}. We outline how we pre-select dominant galaxy candidates. We then detail our approach to modeling probability functions for identifying dominant galaxy candidates based on local stellar mass contrast density measurements. Then we discuss the effectiveness of this method based on the completeness and purity and, finally, we describe our probabilistic method for defining cluster members and hence determining the richness of the galaxy clusters, which is crucial for establishing halo mass-richness relations shown in the end of this section. The analysis was carried out in six photometric redshift intervals divided as follows: [0.1, 0.45[, [0.45, 0.7[, [0.7, 1.05[, [1.05, 1.3[, [1.3, 1.5[, [1.5, 2[. 

Notice that the selection criteria and methods for identifying dominant galaxies and cluster members are applied in a blind manner here, meaning they are implemented independently of the galaxy classifications (Section \ref{sec: definitions}). However, the evaluation and modeling are conducted based on these results, with prior knowledge of these classifications. Additionally, we use perturbed magnitudes (Section \ref{sec: pccones}) as well as photometric redshifts and stellar masses (Section \ref{sec: photoz}).




\subsection{Pre-selection of dominant galaxy candidates}
\label{sec: presel}

The dominant galaxies are the most massive galaxies within a given structure, i.e., galaxy cluster or protocluster. Therefore, it makes sense to use this physical fact to pre-select massive galaxies and thus drastically reduce the number of objects processed in our dominant galaxy detection algorithm (as described in Section \ref{sec: algorithm}). 

The first criterion applied involves a straightforward selection of lightcone objects with photometric stellar masses exceeding a defined limit. To determine this limit, we analyzed the recovered fraction of BCGs as a function of photometric stellar mass cuts across the photometric redshift intervals. From this analysis, we adopted the values $\rm \log(M_{\star, phot}/M_{\odot})$ = [11, 11, 11, 10.5, 10.5, 10.5], increasing with the photometric redshift intervals described above, which ensures recovery fractions of $\rm \gtrsim 90\%$. For redshift intervals above $\rm z = 1$, these limits are lower due to the less accurate stellar mass estimates at higher redshifts, which tend to underestimate the true stellar mass (see Figure \ref{fig: photoz}). This underestimation results in a decrease in the recovered fraction at lower masses.

\begin{figure*}

	\includegraphics[width=\textwidth]{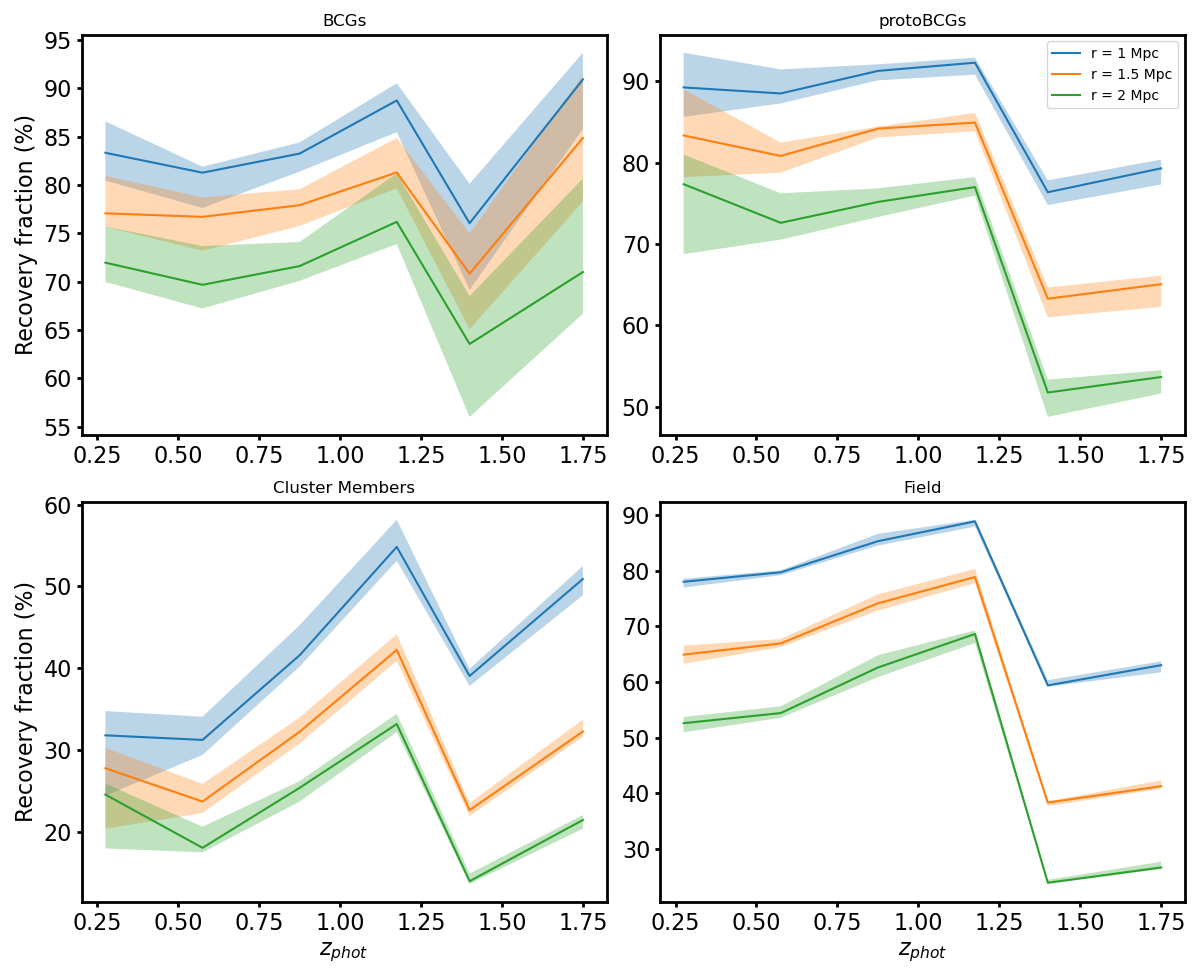}
    \caption{\textit{Recovery fraction} as function of redshift when applying the pre-selection criteria where a given galaxy should be the most massive galaxy in a given cylindrical volume of height $\rm \Delta z$ (Eq. \ref{eq: delta_z}) and transverse radius equal 1, 1.5, or 2 Mpc. The four panels depict the results for BCGs, protoBCGs, Cluster Members, and Field, respectively. The blue, salmon, and green curves (shaded regions), represent the median (in between 16th and 84th percentiles) for the ten lightcones results when using a radius of 1, 1.5, and 2 Mpc, to delimit the cylindrical volume, respectively. The larger error bars at low redshift are due to cosmic variance, while those for BCGs reflect their relative rarity compared to other populations.
}

    \label{fig: pre_sel}
\end{figure*}

Another criterion adopted for the pre-selection of objects was to require that the galaxy be the most massive within a given cylindrical comoving volume around it. We fixed the height of this cylinder as the redshift slice $\rm \Delta z$ (Eq. \ref{eq: delta_z}) and, to choose the radius, we analyzed the recovery fraction of the different types of galaxies (as described in Section \ref{sec: definitions}) after applying this criterion. Our aim here is to retain (remove) the maximum number of dominant (non-dominant) galaxies possible. Figure \ref{fig: pre_sel} illustrates the recovery fraction as a function of the photometric redshift for \textit{BCGs}, \textit{protoBCGs}, \textit{cluster members}, and \textit{field} (refer to Section \ref{sec: definitions}). As this is a pre-selection stage, we made a more conservative choice, opting for a radius of 1 Mpc for all photometric redshift intervals. 

As depicted in the plots presented in Figure \ref{fig: pre_sel}, these choices preserve $\rm \gtrsim$ 80\% of dominant galaxies for all redshift intervals. Dominant galaxies are lost when applying this criterion due to photometric redshift and stellar mass errors. The majority of these lost dominant galaxies lie in lower-mass halos with $\log(M_{\rm{halo}}/M_{\odot}) \lesssim 14.25$ at the redshift in consideration. This criterion proves effective in removing satellite galaxies (\textit{Cluster Members}), where lower recovery fraction values, generally $\rm \lesssim$ 45\%, are observed. This is understandable since satellite galaxies are generally being compared to dominant galaxies in the same structure, which are more massive. Once again, errors in photometric redshifts and stellar mass impact these results.

Regarding field galaxies, this criterion proves inefficient for two main reasons: first, these galaxies are more isolated, making them the most massive within the defined surrounding volume; and second, since these are pre-selected massive galaxies, they might actually be dominant galaxies of groups, i.e., their host halo mass is below $\rm 10^{14} \ M_{\odot}$ and will not reach this threshold at any point in the future simulation. Therefore, these groups do not fit our definition of a cluster or protocluster of galaxies. Notice that this is only a pre-selection criterion. In the next section, we will apply an additional criterion to refine the selection of dominant galaxy candidates based on their associated local density measurements. Since field galaxies tend to reside in lower-density environments, it is likely that they will not be selected in the final stage.

\subsection{Modeling the density contrast distribution}
\label{sec: prob_mod}

For each of the pre-selected dominant galaxy candidates, we calculated the local density contrast following the prescription described in Section \ref{sec: algorithm} (see Eq. \ref{eq: vol_cyl}). Since we have the a priori information about which of the pre-selected galaxies are true dominant and which are not, we can calculate, through the distributions of local density contrasts associated with each galaxy, the probability of a given galaxy being dominant or not, using the following expression:

\begin{equation}
P_{\rm dominant}(\delta \rho_{t}) = \frac{n_{\rm dominant} (\delta \rho > \delta \rho_{t})}{n_{\rm total} (\delta \rho > \delta \rho_{t}) },
    \label{eq: probBCG}
\end{equation}

\noindent
where $n_{\rm dominant} (\delta \rho > \delta \rho_{t})$ is the number of true dominant galaxies with density contrast above a given threshold ($\delta \rho_{ t}$); and $n_{\rm total} (\delta \rho > \delta \rho_{t})$ is the total number of objects above the same threshold.\\

\begin{figure*}

	\includegraphics[width=\textwidth]{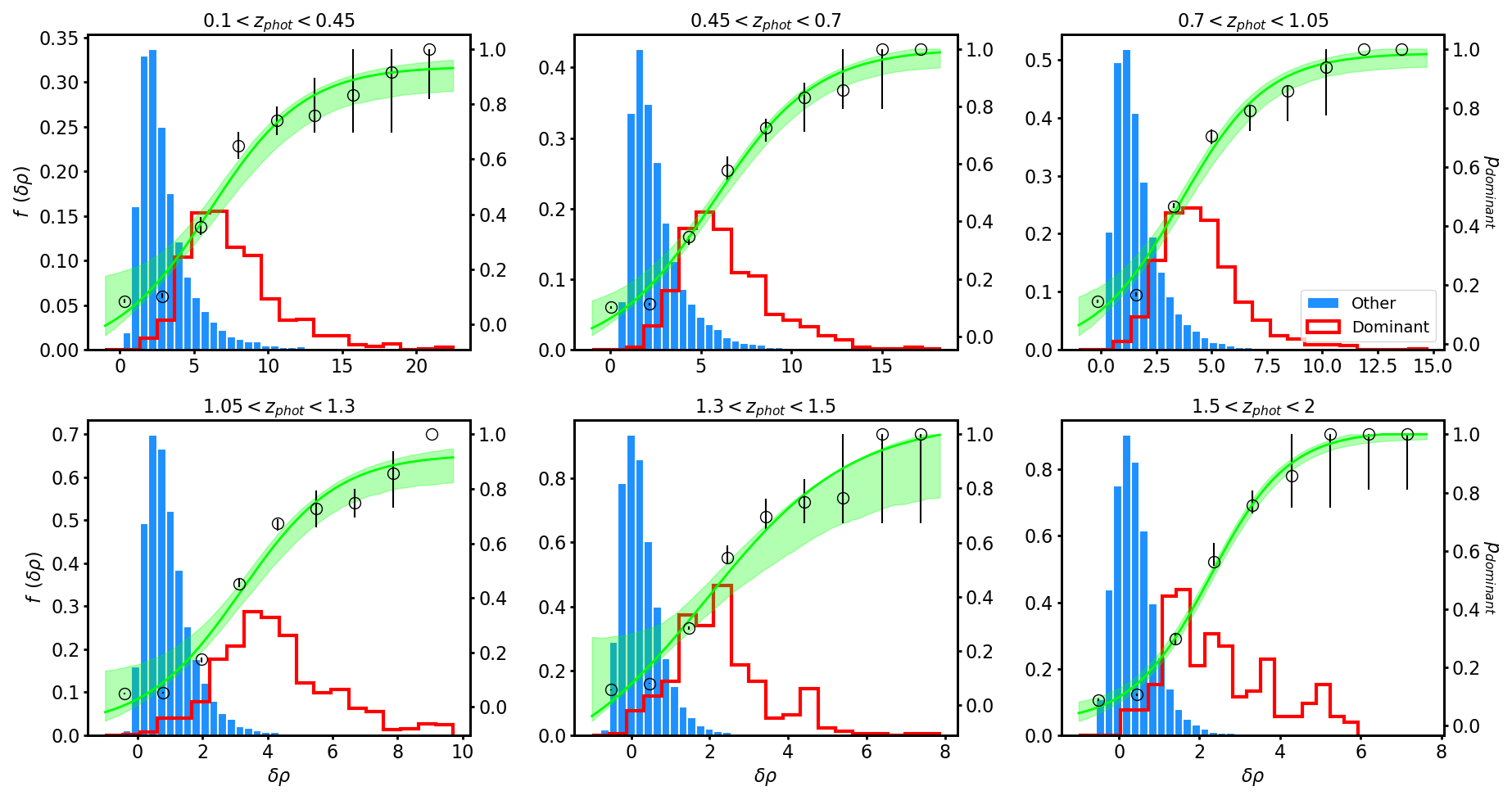}
    \caption{Each panel in this figure represents a redshift interval, as indicated above each plot. The blue histogram depicts the distribution of density contrast for all pre-selected dominant galaxy candidates, while the red histogram represents the distribution only for galaxies classified as BCG or protoBCG, i.e., true dominant galaxies according to our criteria. The histograms have been normalized to unit integral. Black circles with error bars denote the probabilities (y-axis on the right) as a function of density contrast calculated according to equation \ref{eq: probBCG}. The circles represent medians, and the error bars are calculated as the 16 and 84th percentiles from the measurements of the ten mocks. The lime curve represents a fit of a sigmoid function with four free parameters (equation \ref{eq: prob_func}) using median values, while the shaded area delineates fits based on the 16 and 84th percentiles.  
}

    \label{fig: probs}
\end{figure*}

Figure \ref{fig: probs} presents the density contrast distributions and dominant galaxy probability curves for our six redshift intervals for pre-selected galaxies as described in Section \ref{sec: presel}. Probability measurements were conducted mock by mock to quantify the variance of these measurements. As a result, the data points with error bars represent the median $P_{\rm dominant}(\delta \rho_{t})$ and the 16th and 84th percentiles considering the 10 sets of data. We fit these points to a modified sigmoid function with four free parameters ($a$, $b$, $c$, and $d$):

\begin{equation}
    f(\delta \rho; a, b, c, d) = \frac{a}{\{1 + exp[-b(\delta \rho - c)]\} + d}.
    \label{eq: prob_func}
\end{equation}

\noindent
The lime-colored curves are fits based on the median of the data points and the shaded area delineates the region between the fits using the 16th and 84th percentiles. Table \ref{tab: prob_pars} shows the parameters obtained for these parameters when using the median contrast density measurements.

\begin{table*}
\centering
\caption{dominant galaxies probability function parameters (Eq. \ref{eq: prob_func}) for six redshift intervals.}
\begin{tabular}{ccccc}
\hline
Redshift bin & a & b & c & d \\
\hline
0.10 -- 0.45 & 1.045 $\pm$ 0.281 & 0.311 $\pm$ 0.130 & 6.058 $\pm$ 1.760 & -0.109 $\pm$ 0.240 \\
0.45 -- 0.70 & 1.061 $\pm$ 0.180 & 0.354 $\pm$ 0.100 & 5.707 $\pm$ 0.900 & -0.061 $\pm$ 0.144 \\
0.70 -- 1.05 & 1.009 $\pm$ 0.181 & 0.511 $\pm$ 0.150 & 3.594 $\pm$ 0.760 & -0.024 $\pm$ 0.157 \\
1.05 -- 1.30 & 1.002 $\pm$ 0.215 & 0.675 $\pm$ 0.272 & 3.194 $\pm$ 0.604 & -0.075 $\pm$ 0.160 \\
1.30 -- 1.50 & 1.315 $\pm$ 0.582 & 0.540 $\pm$ 0.307 & 1.924 $\pm$ 1.217 & -0.266 $\pm$ 0.460 \\
1.50 -- 2.00 & 1.002 $\pm$ 0.069 & 1.030 $\pm$ 0.155 & 2.261 $\pm$ 0.158 & 0.101 $\pm$ 0.054 \\

\hline
\end{tabular}
\label{tab: prob_pars}
\end{table*}

The distribution of $\delta \rho$ corresponding to true dominant galaxies skews toward regions of higher density contrast than all other galaxies and, consequently, higher probability. However, there is a considerable overlap between the two histograms in all redshift bins. Except for the highest redshift bin, the density contrast distributions of dominant galaxies exhibits a long tail towards higher contrasts. The peaks of the red histograms are around $\delta \rho \sim 8$ for $\rm z < 1.05$, and shift to $\delta \rho \sim 5$ for $\rm 1.05 < z < 1.5$. In the case of $\rm z > 1.5$, the histogram is noisy, but it is evident that objects are concentrated in regions with lower density contrast than at lower redshift bins, with a peak around $\delta \rho \sim 2$. 

\subsection{Evaluating the Efficiency of (Proto)BCG Detection}
\label{sec: effec}

Employing the probability models, we computed the likelihood that each candidate dominant galaxy truly is a dominant galaxy. The quality of the results can be assessed using a combination of completeness and purity. Both metrics are defined based on a specified probability threshold, which, when applied, allows the selection of a sample with a given completeness and purity percentage. 

For a given probability threshold \rm $P_{t}$ and a given redshift bin, completeness is defined as the ratio of the number of true dominant galaxies with a probability higher than the threshold $P_{\rm{dominant}}(\delta \rho) > P_{t}(\delta \rho)$ to the total number of true dominant galaxies. Purity, on the other hand, represents the ratio of the number of selected true dominant galaxies with $P_{\rm{dominant}}(\delta \rho) > P_{t}(\delta \rho)$ to the total number of selected galaxies with the same cutoff. Thus, in an ideal scenario, one would determine a probability threshold where all dominant galaxies are preserved (100\% completeness), and all other galaxy populations are removed (100\% purity).

\begin{figure*}

	\includegraphics[width=\textwidth]{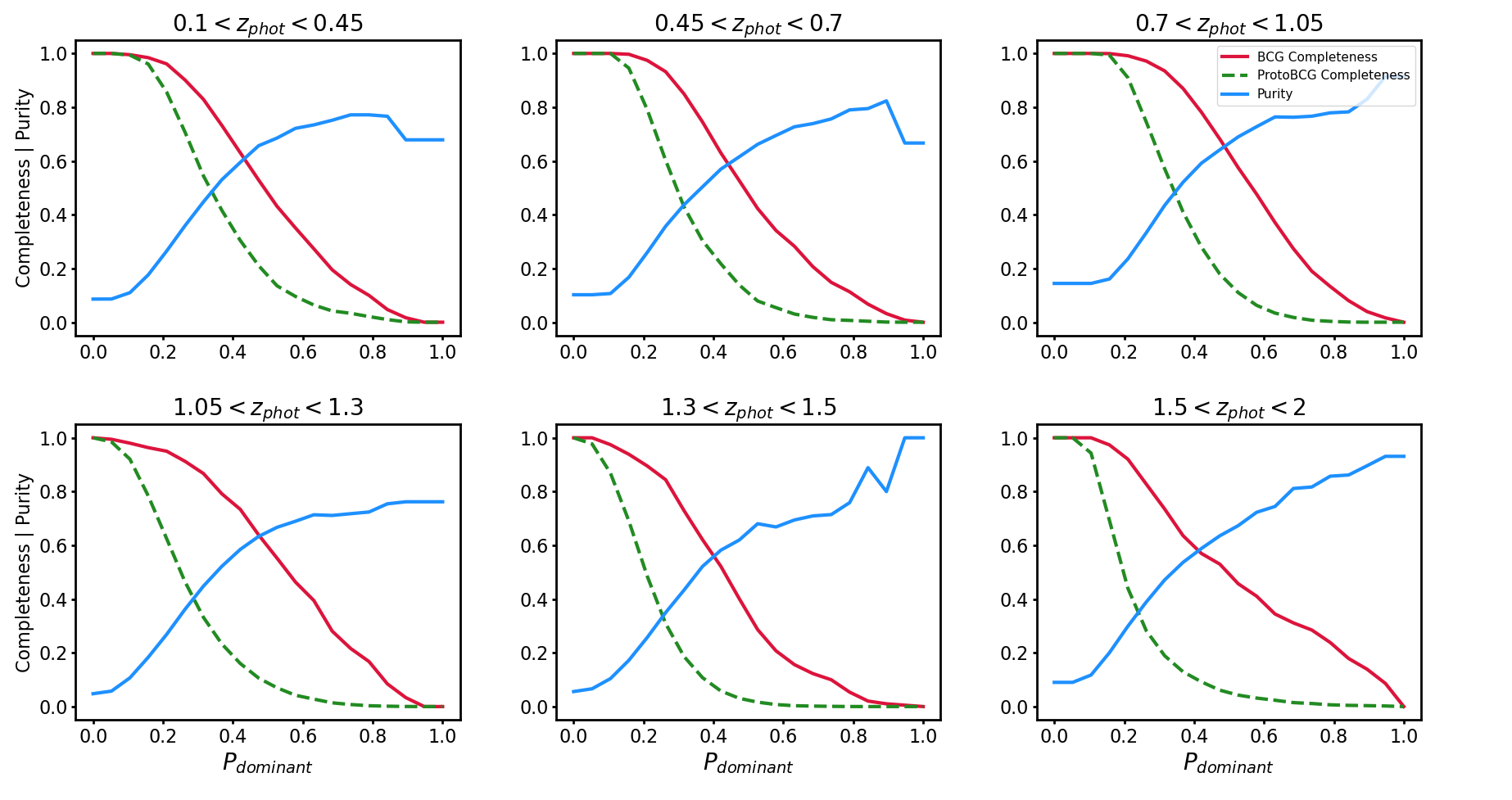}
    \caption{Each panel in this figure represents a redshift interval, as indicated above each plot. Completeness and purity are shown as a function of the probability of galaxies being dominant. The solid red curve and the dashed green curve denote completeness for BCGs and protoBCGs, respectively. Meanwhile, the blue curve represents purity considering both BCGs and protoBCGs. At high probability thresholds, the number of objects becomes small not only in the numerator but also in the denominator when computing purity, leading to increased statistical fluctuations. This effect is particularly noticeable in the $1.3 < z < 1.5$ panel, but the noise is similar in other redshift intervals for $P_{\rm{dominant}} \gtrsim 0.7$.
}

    \label{fig: comp_pur}
\end{figure*}

Taking into account these quantities, Figure \ref{fig: comp_pur} depicts the completeness curve (solid red for BCGs and dashed green for protoBCGs) and the purity curve (blue) as a function of $P_{\rm{dominant}}$ for each redshift interval. The purity calculation takes into account dominant galaxies in general, i.e., both BCGs and protoBCGs. 

\begin{figure*}

	\includegraphics[width=\textwidth]{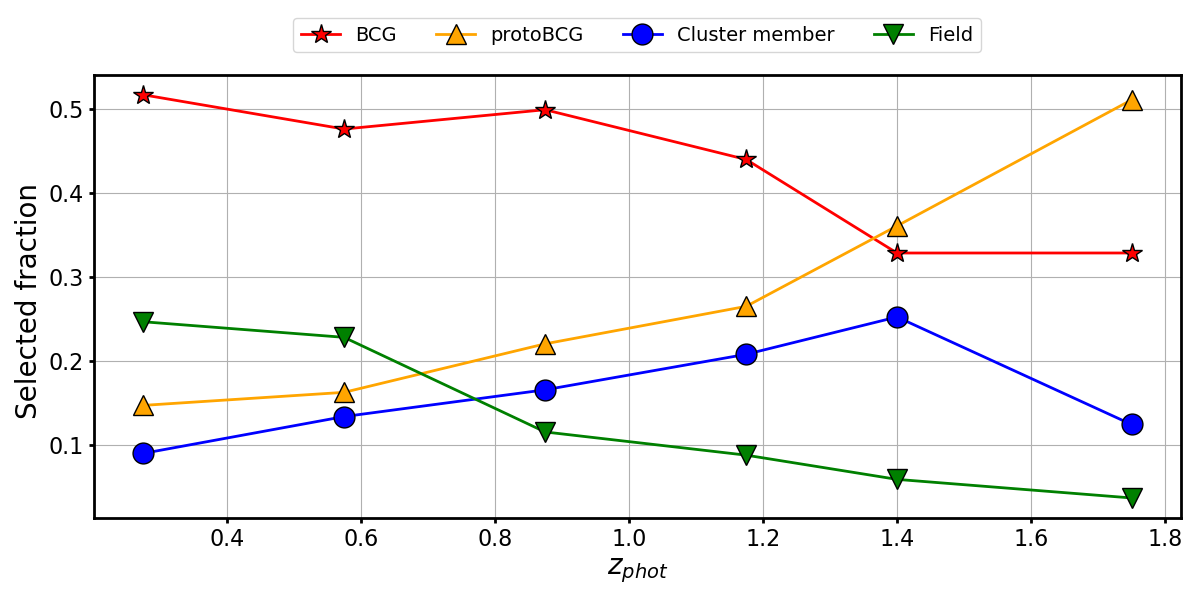}
    \caption{Fraction of candidate BCG`s selected of each true galaxy type when using the expected number of BCGs in each redshift interval with the highest probabilities as a function of redshift. The red stars, orange triangles, blue circles, and green inverted triangles refer to \textit{BCGs}, \textit{protoBCGs}, \textit{Cluster Members}, and \textit{Field} galaxies, respectively. The false positive rate is the sum of the blue and green curves.  
}
    \label{fig: sel_objs}
\end{figure*}

Since we have the information of the expected number of BCGs in each redshift interval $\rm n(BCG|\Delta z)$, i.e., the number of true BCGs pre-selected as described in Section \ref{sec: presel}, we can use this value as a reference to assess the percentages of different object types selected this way. Thus, a number of objects equal to $\rm n(BCG|\Delta z)$ with the highest probabilities are selected. Figure \ref{fig: sel_objs} illustrates the percentages of selected object types that we defined for this work (see Section \ref{sec: definitions}) when choosing the top $\rm n(BCG|\Delta z)$ objects with the highest probabilities as a function of photometric redshift. The number of objects selected this way for each redshift interval considering our ten mocks with $\rm 36 \ deg^{2}$, are: 1083, 1281, 1971, 857, 341, and 137. We can select this number of clusters with a threshold $P_{\rm{dominant}} \gtrsim 46, 50, 57, 61, 60, 79\%$.

\begin{figure}
    \centering
	\includegraphics[width=\columnwidth]{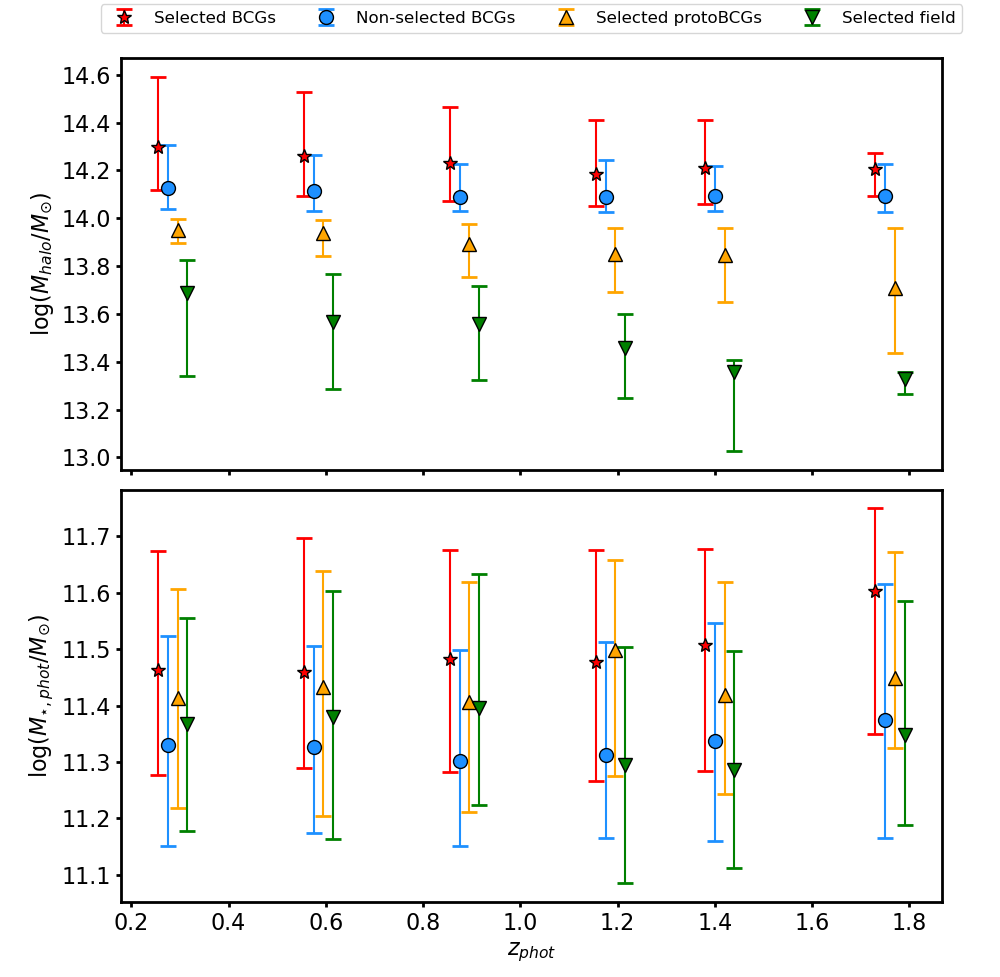}
    \caption{The top (bottom) panel displays the halo (photometric stellar) mass of selected galaxies as a function of redshift. Red stars, orange triangles, and green inverted triangles denote median values for BCGs, protoBCGs, and field galaxies that were selected with the highest probabilities based on the expected number of BCGs in a given redshift bin. Blue points represent true BCGs which were not selected. A slight horizontal shift was applied to the different markers to improve visualization.
}

    \label{fig: BCG_sel_not_sel}
\end{figure}

The percentage of true BCGs smoothly decreases from 52\% to 44\% up to the redshift interval $\rm 1.05 < z < 1.3$. Afterward, this value decreases more rapidly, reaching 33\% in the last redshift interval. This behavior is partly a consequence of the quality of photometric redshift and stellar mass estimates. Beyond $\rm z > 1.3$, the spectral coverage of the HSC-SSP bands no longer includes important features, such as the 4000 \angstrom \xspace break, and the fraction of objects with photometry in the 3.6 and 4.5 $\mu$m bands also decreases considerably. The consequence is a deterioration in the quality of stellar mass and photometric redshift estimates which are dependent on the object photometry. Figure \ref{fig: BCG_sel_not_sel} shows that in all redshift intervals, the selected BCGs are consistently the most massive galaxies. Additionally, selected protoBCGs tend to be more massive than non-selected BCGs. At $z > 1.3$, the selected BCGs become increasingly massive, which can be attributed to a natural bias toward selecting higher-mass objects at higher redshifts. However, in this regime, protoBCGs become much more abundant (see Figure \ref{fig: n_cl_prot}), and consequently, a larger fraction of them is also selected (Figure \ref{fig: sel_objs}).

ProtoBCG selection increases smoothly with z from 15\% up to 27\% at $\rm 1.05 < z < 1.3$ and more steeply in the last two intervals, reaching 51\% of the objects selected. These galaxies are slightly more massive than BCGs that were not selected by our criteria, and they also inhabit relatively massive halos with $13.8 \lesssim \log(M_{\rm{halo}}/M_{\odot}) < 14$ (Figure \ref{fig: BCG_sel_not_sel}). With the significant increase in protoclusters with redshift, protoBCGs become dominant in our selection at z $>$ 1.3.

The contamination by other cluster galaxies (blue points in Figure \ref{fig: sel_objs}) increases steadily until $\rm z = 1.5$ and, in the last interval, decreases drastically. This behavior can be explained due to our pre-selection criterion of choosing the most massive galaxy in a given volume, as described in Section \ref{sec: presel}. The $\textit{Recovery fraction}$ of this type of galaxy increases with redshift because cluster members are less concentrated in the central regions of the structure, which increases the number of cases in which satellite galaxies are located at a distance from the BCG greater than the maximum radius. In the case of $\rm 1.5 < z < 2$, protoBCGs start to dominate the selection of massive galaxies inhabiting overdense regions.

Objects classified as \textit{field} galaxies decrease with redshift. At first glance, this behavior is not clear, as one might expect this type of contamination to increase with redshift due to the poorer photometry and greater projection effects at high-z. However, field galaxies selected according to our criteria are massive galaxies that inhabit relatively massive halos, with $log(M_{\rm{halo}}/M_{\odot}) \gtrsim 13.5$ (see Figure \ref{fig: BCG_sel_not_sel}), and can be interpreted as potential dominant galaxies of massive groups, but will not reach the mass threshold to be classified as protoclusters. The contamination of these galaxies is higher at low redshifts since there are more massive groups that will not have time to reach the cluster mass threshold by $\rm z = 0$.

\subsection{Evaluating the Efficiency of (Proto)cluster Detection}
\label{sec: effec_cl}

In this section, we evaluate the completeness and purity of our galaxy (proto)cluster selection. The primary distinction compared to Section \ref{sec: effec}, where we assessed the detection of dominant galaxies, is that here we consider contamination exclusively when selecting a galaxy within a halo that does not meet our definition of a galaxy (proto)cluster (see Section \ref{sec: definitions}), i.e., a field galaxy.

\begin{figure*}
    \centering
	\includegraphics[width=\textwidth]{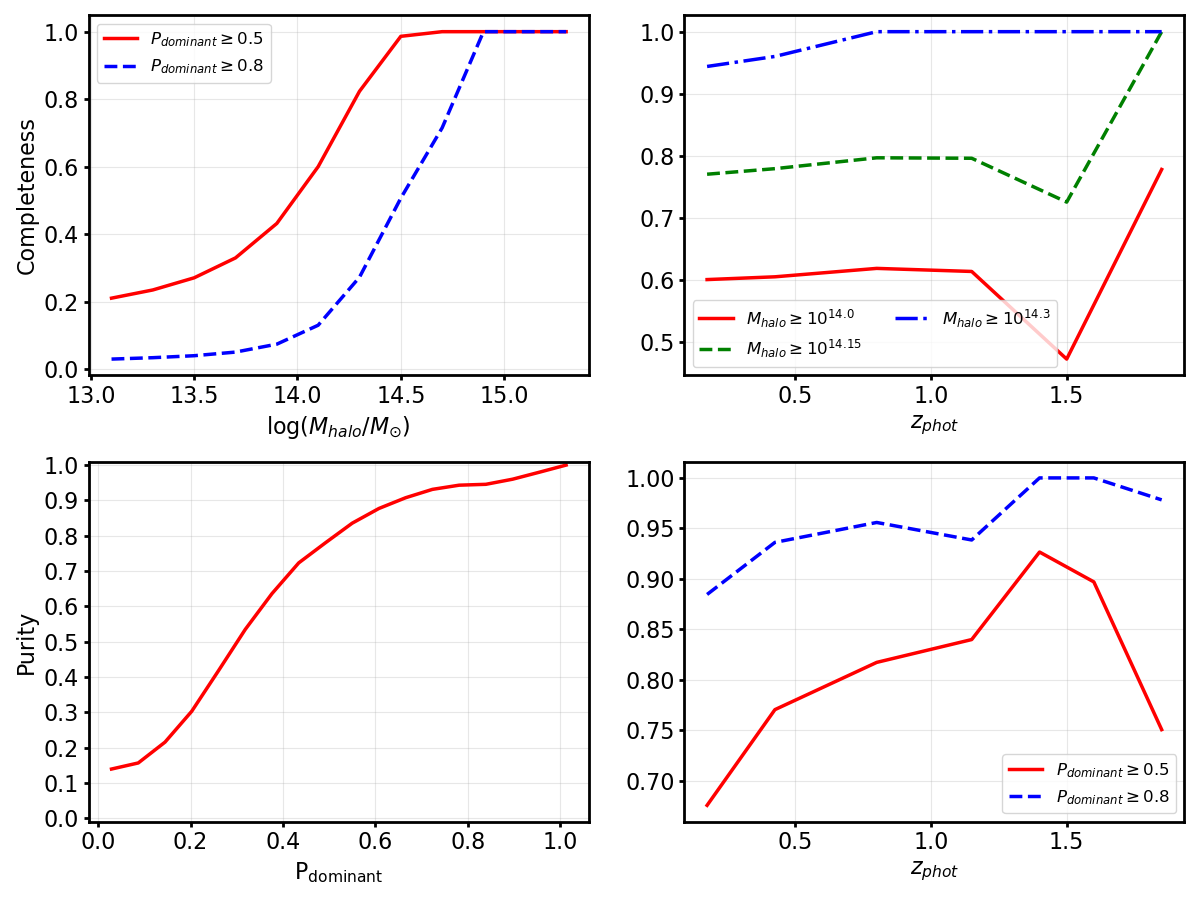}
    \caption{Completeness and purity of (proto)cluster detection as a function of various quantities. Top-left panel: Completeness as a function of halo mass for two selection thresholds of dominant galaxy probability, $P_{\rm dominant} \geq 0.5$ (solid red) and $P_{\rm dominant} \geq 0.8$ (dashed blue). Top-right panel: Completeness as a function of photometric redshift, considering three different minimum halo mass thresholds: $M_{\rm halo} \geq 10^{14.0} \ M_{\odot}$ (solid red), $M_{\rm halo} \geq 10^{14.15} \ M_{\odot}$ (dashed green), and $M_{\rm halo} \geq 10^{14.3} \ M_{\odot}$ (dash-dotted blue). Bottom-left panel: Purity as a function of the probability threshold for dominant galaxy selection. Bottom-right panel: Purity as a function of photometric redshift for two probability thresholds, $P_{\rm dominant} \geq 0.5$ (solid red) and $P_{\rm dominant} \geq 0.8$ (dashed blue).
}

    \label{fig: protocl_comp_pur}
\end{figure*}

Figure \ref{fig: protocl_comp_pur} displays four panels illustrating completeness (top row) and purity (bottom row) fractions as functions of various quantities. In the top-left panel, completeness is shown as a function of the halo mass of the structures. The curves represent results for different cuts in the probability, $P_{\rm{dominant}}$, as indicated in the legend. As expected, the selection becomes more complete for more massive structures, reaching nearly $\sim 100\%$ for structures with $\log(M_{\rm{halo}}/M_{\odot}) \gtrsim 14.5$ when selecting dominant galaxies with $P_{\rm{dominant}} \geq 0.5$.

In the top-right panel, completeness is shown as a function of photometric redshift. The results are divided based on three different lower limits of $M_{\rm{halo}}$ for structures selected with $P_{\rm{dominant}} \geq 0.5$. Notably, completeness increases significantly when the halo mass threshold is slightly raised. The selection is nearly complete across the entire redshift range for structures with $log(M_{\rm{halo}}/M_{\odot}) \gtrsim 14.3$. The observed increase in completeness with redshift is primarily due to the decreasing number of cluster-mass structures at higher redshifts. Those that do exist tend to reside in the highest-density peaks and have already reached the cluster mass threshold, continuing to grow and eventually becoming the most massive clusters in the local Universe. As a result, when applying our algorithm at high redshifts, a larger fraction of these rare, well-formed structures is successfully selected, leading to higher completeness values.

As expected, purity increases with $P_{\rm{dominant}}$, as shown in the bottom-left panel. It can be observed that a sample selected with $P_{\rm{dominant}} \geq 0.5$ is approximately $80\%$ pure, while for $P_{\rm{dominant}} \geq 0.8$, the purity rises to around $95\%$. The bottom-right panel illustrates how these purity levels vary with photometric redshift. Interestingly, there is an increase in purity at higher redshifts. This effect, as discussed in Section \ref{sec: effec}, arises due to greater contamination from massive field galaxies—those that do not fit our definition of (proto)clusters. At lower redshifts, halos will not have time to evolve and surpass the mass threshold we use to define galaxy clusters. Conversely, at higher redshifts, massive groups are more likely to surpass this threshold over time, allowing us to classify them as protoclusters, which are not considered contaminants in this analysis.

\subsection{(Proto)cluster members}
\label{sec: cm_sel}

To determine which galaxies are members of a (proto)cluster, we analyzed, for each pre-selected dominant galaxy, the distributions of physical properties of galaxies which are in the same redshift slice (Eq. \ref{eq: delta_z}) and at a maximum angular diameter transverse distance of 1 Mpc from the dominant galaxy. For this, we identified three samples: one formed by galaxies that are indeed (proto)cluster members, and the other two formed by contamination from foreground or background objects. 

\begin{figure*}
    \centering
	\includegraphics[width=\textwidth]{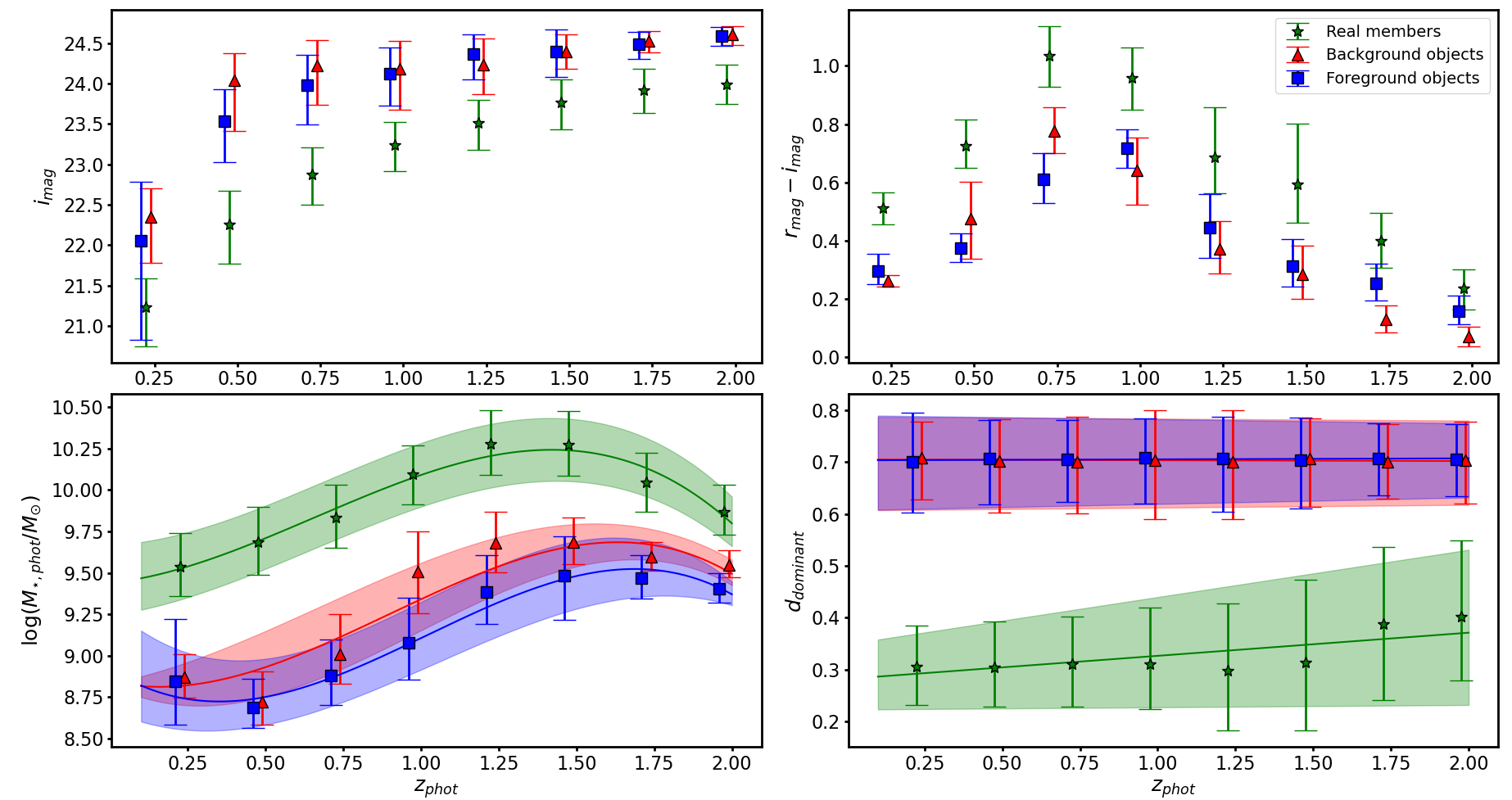}
    \caption{i-band magnitude, r-i color, stellar mass, and transverse angular diameter distance from the dominant galaxy ($\rm d_{dominant}$) as a function of photometric redshift. The markers denote the medians in redshift bins, and the bars are delimited by the 16th and 84th percentiles. Green, blue, and red colors stand for real (proto)cluster members, foreground, and background galaxies, respectively. A slight horizontal shift was applied to the different markers to improve visualization. 
}

    \label{fig: members}
\end{figure*}

Figure \ref{fig: members} shows four plots with i-band magnitude, r-i observed color, photometric stellar mass, and distance from the dominant galaxy ($\rm d_{dominant}$) of these three samples as a function of photometric redshift. The markers denote the medians of the properties in photometric redshift bins, and the bars are delimited by the 16th and 84th percentiles. Green, blue, and red colors stand for real (proto)cluster members, foreground, and background galaxies, respectively. The idea of these plots is to determine which properties best distinguish real members from contamination.

(Proto)cluster members are brighter, redder, more massive, and are located close to the BCG. Based on these plots, we choose photometric stellar mass and the distance from the BCG to select (proto)cluster members. The lines on these plots represent fits considering the median at each redshift bin for each sample, while the shaded areas delineate the region between fitted functions using the 16th and 84th percentiles. From these fits, we calculate the probability of a given object being a true member $\rm P(Member|M_{\star, phot}, d_{dominant}, z_{dominant})$ or contamination $\rm P(Cont|M_{\star, phot}, d_{dominant}, z_{dominant})$. To do this, we assumed that the objects are distributed following a Normal distribution with a mean equal to the value calculated by the fitted function and a standard deviation calculated for each redshift bin. Finally, the galaxy is considered a member of the structure if $\rm P(Member|M_{\star, phot}, d_{dominant}, z_{dominant}) > P(\rm{Cont}|M_{\star, phot}, d_{dominant}, z_{dominant})$ and the richness ($\lambda$) of the (proto)cluster is thus defined as the number of galaxies which satisfy this condition.

\begin{figure*}
    \centering
	\includegraphics[width=\textwidth]{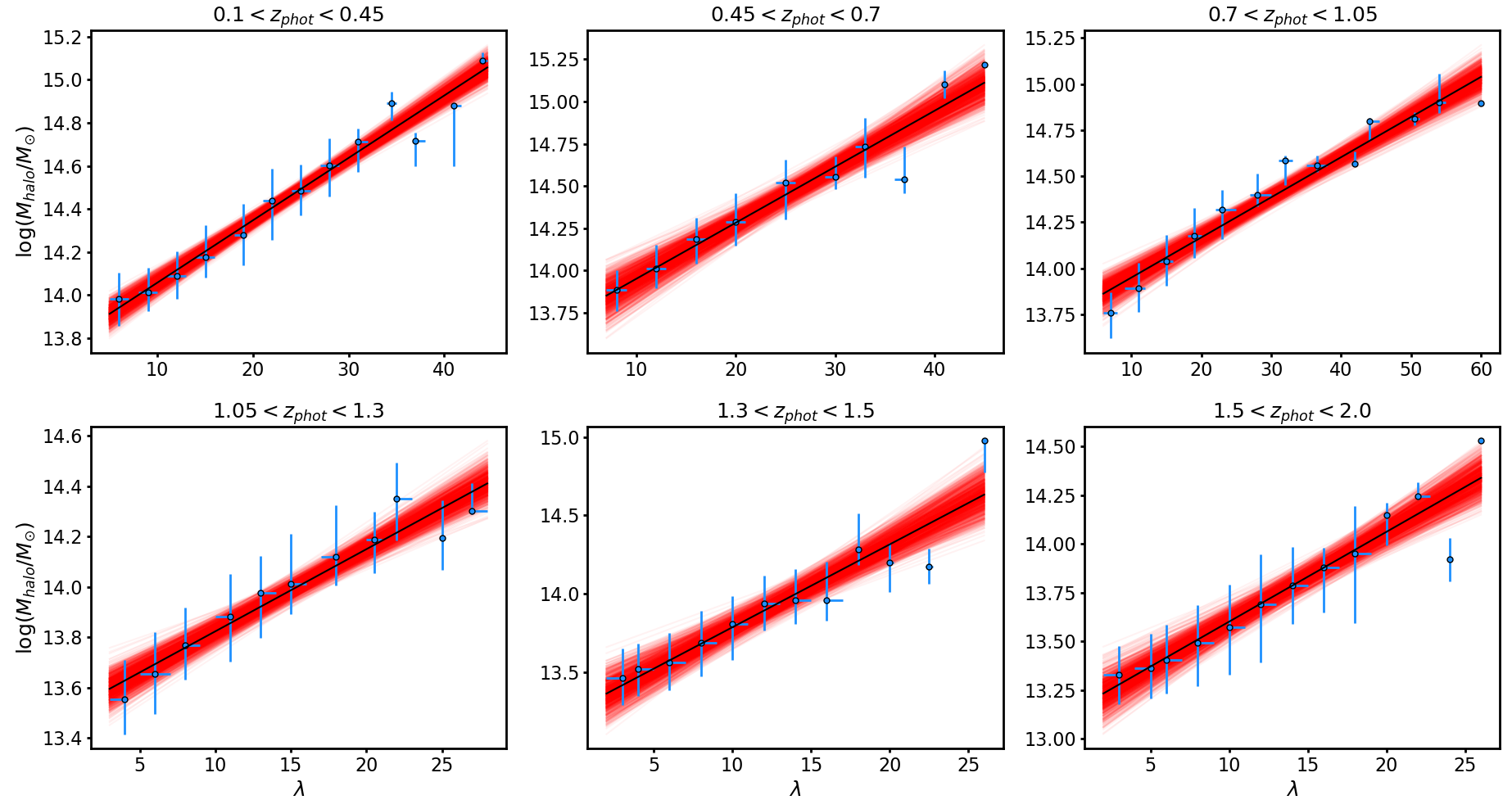}
    \caption{Halo mass-richness relations for different redshift intervals, as indicated above each panel. The Y-axis represents the true halo masses from the Millennium simulation. The red shaded area corresponds to bootstrap resampling of the fits, where the estimated parameters are varied within their respective uncertainties, illustrating the statistical dispersion of the relation. 
}

    \label{fig: mass_rich}
\end{figure*}

Finally, since we have information about the mass of the structure in the mocks, we obtained halo mass-richness relations (Figure \ref{fig: mass_rich}), which will be useful to estimate the (proto)cluster halo masses when we apply the same selection criteria in the HSC-SSP Wide Survey. Table \ref{tab: mass_rich_pars} shows the slope ($\alpha$) and the intercept ($\beta$) for each redshift interval.

The halo mass-richness relations for protoclusters were better fitted using a log-linear relation, $\log(M_{\rm halo})$–$\lambda$, rather than the commonly observed log-log relations. When analyzing structures from the pure simulation—i.e., the halo mass and the total number of members—log-log space indeed provides a better fit. However, when observational constraints are incorporated, particularly photometric redshifts, the results favor a log-linear relation. This highlights the impact of observational uncertainties on the derived scaling relations and emphasizes the necessity of adapting models to account for these effects.

\begin{table}
\centering
\caption{Slope ($\alpha$) and intercept ($\beta$) for halo mass-richness relations at different redshift intervals.}
\begin{tabular}{ccc}
\hline
Redshift bin & $\alpha$ & $\beta$ \\
\hline
0.10 -- 0.45 & 0.029 $\pm$ 0.002 & 13.769 $\pm$ 0.042 \\
0.45 -- 0.70 & 0.033 $\pm$ 0.003 & 13.620 $\pm$ 0.098 \\
0.70 -- 1.05 & 0.022 $\pm$ 0.002 & 13.732 $\pm$ 0.058 \\
1.05 -- 1.30 & 0.033 $\pm$ 0.003 & 13.498 $\pm$ 0.053 \\
1.30 -- 1.50 & 0.053 $\pm$ 0.006 & 13.258 $\pm$ 0.095 \\
1.50 -- 2.00 & 0.046 $\pm$ 0.005 & 13.140 $\pm$ 0.074 \\

\hline
\end{tabular}
\label{tab: mass_rich_pars}
\end{table}

\section{Summary}
\label{sec: summary}

This paper represents the first part of an intended two-part series. We propose a novel method for identifying galaxy clusters and protoclusters at $\rm 0.1 < z < 2$ range in the the HSC-SSP wide optical photometric survey, with the identification of the dominant galaxy as the first step, i.e., BCGs or protoBCGs. 

Our approach involves extensive use of simulated data that emulate observations from the HSC-SSP, via lightcones called PCcones. This paper presents the algorithm (Section \ref{sec: algorithm}), how we prepared PCcones data to apply the algorithm (Section \ref{sec: mock_prepare}), and the results of this application (Section \ref{sec: application}). The PCcones mocks were constructed by applying semi-analytical models of galaxy formation and evolution (\texttt{L-GALAXIES}) to Millennium Simulation data (Section \ref{sec: pccones}). It provides an observational perspective by constructing lightcones and obtaining magnitudes for galaxies adopting transmission filters similar to those used in observations, allowing for the inclusion of the completeness limit of the HSC-SSP Wide Survey, for example. 

We used data from the HSC-SSP Wide Survey to model errors in magnitudes for the different filters and perturb these magnitudes according to these models (Section \ref{sec: pccones}). With the perturbed magnitudes, we estimate photometric stellar mass and redshifts for all galaxies in the mock dataset (Section \ref{sec: photoz}).

Another crucial aspect is that PCcones retain information about which galaxies belong to structures with information about the dark matter halo mass. This allows us to use definitions based on this mass to define galaxy clusters, and thus which galaxies are part of them. Here, we consider galaxy clusters as structures with $M_{\rm{halo}} \geq 10^{14} \/\ M_{\odot}$, at the observed redshift. Additionally, information is available about which halos will exceed this mass limit at some future point, i.e., $\rm 0 < z < z_{obs}$, which we define as protoclusters (Section \ref{sec: definitions}). 

Our strategy involves using this dataset to optimize selection criteria (Section \ref{sec: presel}) and obtain probabilistic models to define dominant galaxies of clusters or protoclusters, based on local stellar mass contrast density associated with pre-selected massive galaxies (Sections \ref{sec: presel} and \ref{sec: prob_mod}) and galaxy (proto)cluster members, based on stellar mass and the distance to the BCG of real (proto)cluster members, enabling us to establish halo mass-richness relations for different redshift intervals (Section \ref{sec: cm_sel}).

Our results demonstrate that it is possible to obtain a sample of dominant galaxy candidates with $\gtrsim$ 65\% purity by selecting pre-selected massive galaxies according to our criteria and with a probability of being dominant $\rm P_{\rm dominant} > 50\%$ (Figure \ref{fig: comp_pur}). Most of the contamination at $\rm z_{phot} \lesssim 0.7$ (about 20\%) is due to massive galaxies $\rm \log(M_{\star, phot}/M_{\odot}) \gtrsim 11.3$ residing in relatively massive halos $13.5 \lesssim \log(M_{\rm{halo}}/M_{\odot}) < 14$ (Figure \ref{fig: BCG_sel_not_sel}). According to our criteria, such galaxies do not inhabit clusters or protoclusters and thus were classified as \textit{Field}. At higher redshifts, the main source of contamination comes from other massive satellite galaxies, classified as \textit{cluster member} (Figure \ref{fig: BCG_sel_not_sel}).

Considering a selection of galaxy (proto)clusters with $\rm P_{\rm dominant} > 50\%$—regardless of whether the dominant galaxy is correctly identified—the sample achieves 80\% purity and 50\% completeness for structures with $M_{\rm{halo}} \geq 10^{14} \ M_{\odot}$, reaching 100\% completeness for $M_{\rm{halo}} \geq 10^{14.5} \ M_{\odot}$.

Finally, using the pre-selected dominant galaxies (Section \ref{sec: presel}), we fit the photometric stellar mass and the distance from the dominant galaxy as a function of photometric redshift both for member galaxies of the structures, and contamination due to other foreground or background galaxies (Figure \ref{fig: members}). From these fits, we defined the richness of the structure based on the number of galaxies with a higher probability of being true members than contamination and we obtained halo mass-richness relations for different redshift intervals (Figure \ref{fig: mass_rich}).

In the second paper in this series, we will apply this algorithm with the selection criteria and probabilistic models for the selection of dominant and satellite galaxies to the photometric data of the HSC-SSP Wide Survey. Additionally, we will compare our findings with other cluster finder algorithms previously applied to the HSC data and with galaxy clusters identified by X-ray emission.

The methods presented in this work can be adapted to other existing and upcoming multi-band photometric surveys--such as the Southern Photometric Local Universe Survey \citep[S-PLUS; ][]{mendesdeoliveira19}, the Dark Energy Survey \citep[DES; ][]{abbott21}, and the Rubin Observatory Legacy Survey of Space and Time \citep[LSST; ][]{ivezic19}--by incorporating their specific observational constraints in the mocks accordingly to redefine selection criteria, adjust the modeling of dominant galaxy identification and cluster membership assignment.


\begin{acknowledgments}
\section*{Acknowledgements}
MCV acknowledges the Fundaç\~ao de Amparo à Pesquisa do Estado de S\~ao Paulo (FAPESP; 2021/06590-0) for supporting his PhD and Research Internship Abroad at the Department of Astrophysical Sciences, Princeton University. He also thanks the Department of Astrophysical Sciences at Princeton University for its financial support in making this internship possible. PA-A thanks the Coordenaç\~ao de Aperfeiçoamento de Pessoal de Nível Superior – Brasil (CAPES), for supporting his PhD scholarship (project 88882.332909/2020-01). LSJ acknowledges the support from CNPq (308994/2021-3) and FAPESP (2011/51680-6). 

\end{acknowledgments}

\vspace{5mm}

\software{Numpy \citep{harris2020}, Pandas \citep{reback2020}, Scipy \citep{scipy}, Matplotlib \citep{hunter2007}, Astropy \citep{astropy}, Tensorflow \citep{tensorflow2015}, Keras \citep{chollet18}, Sklearn \citep{pedregosa2011}}


\begin{thebibliography}{}
\expandafter\ifx\csname natexlab\endcsname\relax\def\natexlab#1{#1}\fi
\providecommand{\url}[1]{\href{#1}{#1}}
\providecommand{\dodoi}[1]{doi:~\href{http://doi.org/#1}{\nolinkurl{#1}}}
\providecommand{\doeprint}[1]{\href{http://ascl.net/#1}{\nolinkurl{http://ascl.net/#1}}}
\providecommand{\doarXiv}[1]{\href{https://arxiv.org/abs/#1}{\nolinkurl{https://arxiv.org/abs/#1}}}

\bibitem[{Abadi {et~al.}(2015)Abadi, Agarwal, Barham, Brevdo, Chen, Citro, Corrado, Davis, Dean, Devin, Ghemawat, Goodfellow, Harp, Irving, Isard, Jia, Jozefowicz, Kaiser, Kudlur, Levenberg, Man\'{e}, Monga, Moore, Murray, Olah, Schuster, Shlens, Steiner, Sutskever, Talwar, Tucker, Vanhoucke, Vasudevan, Vi\'{e}gas, Vinyals, Warden, Wattenberg, Wicke, Yu, \& Zheng}]{tensorflow2015}
Abadi, M., Agarwal, A., Barham, P., {et~al.} 2015, {TensorFlow}: Large-Scale Machine Learning on Heterogeneous Systems.
\newblock \url{https://www.tensorflow.org/}

\bibitem[{{Abbott} {et~al.}(2021){Abbott}, {Adam{\'o}w}, {Aguena}, {Allam}, {Amon}, {Annis}, {Avila}, {Bacon}, {Banerji}, {Bechtol}, {Becker}, {Bernstein}, {Bertin}, {Bhargava}, {Bridle}, {Brooks}, {Burke}, {Carnero Rosell}, {Carrasco Kind}, {Carretero}, {Castander}, {Cawthon}, {Chang}, {Choi}, {Conselice}, {Costanzi}, {Crocce}, {da Costa}, {Davis}, {De Vicente}, {DeRose}, {Desai}, {Diehl}, {Dietrich}, {Drlica-Wagner}, {Eckert}, {Elvin-Poole}, {Everett}, {Evrard}, {Ferrero}, {Fert{\'e}}, {Flaugher}, {Fosalba}, {Friedel}, {Frieman}, {Garc{\'\i}a-Bellido}, {Gaztanaga}, {Gelman}, {Gerdes}, {Giannantonio}, {Gill}, {Gruen}, {Gruendl}, {Gschwend}, {Gutierrez}, {Hartley}, {Hinton}, {Hollowood}, {Honscheid}, {Huterer}, {James}, {Jeltema}, {Johnson}, {Kent}, {Kron}, {Kuehn}, {Kuropatkin}, {Lahav}, {Li}, {Lidman}, {Lin}, {MacCrann}, {Maia}, {Manning}, {Maloney}, {March}, {Marshall}, {Martini}, {Melchior}, {Menanteau}, {Miquel}, {Morgan}, {Myles}, {Neilsen}, {Ogando}, {Palmese}, {Paz-Chinch{\'o}n}, {Petravick},
  {Pieres}, {Plazas}, {Pond}, {Rodriguez-Monroy}, {Romer}, {Roodman}, {Rykoff}, {Sako}, {Sanchez}, {Santiago}, {Scarpine}, {Serrano}, {Sevilla-Noarbe}, {Smith}, {Smith}, {Soares-Santos}, {Suchyta}, {Swanson}, {Tarle}, {Thomas}, {To}, {Tremblay}, {Troxel}, {Tucker}, {Turner}, {Varga}, {Walker}, {Wechsler}, {Weller}, {Wester}, {Wilkinson}, {Yanny}, {Zhang}, {Nikutta}, {Fitzpatrick}, {Jacques}, {Scott}, {Olsen}, {Huang}, {Herrera}, {Juneau}, {Nidever}, {Weaver}, {Adean}, {Correia}, {de Freitas}, {Freitas}, {Singulani}, {Vila-Verde}, \& {Linea Science Server}}]{abbott21}
{Abbott}, T.~M.~C., {Adam{\'o}w}, M., {Aguena}, M., {et~al.} 2021, \apjs, 255, 20, \dodoi{10.3847/1538-4365/ac00b3}

\bibitem[{{Aguena} {et~al.}(2021){Aguena}, {Benoist}, {da Costa}, {Ogando}, {Gschwend}, {Sampaio-Santos}, {Lima}, {Maia}, {Allam}, {Avila}, {Bacon}, {Bertin}, {Bhargava}, {Brooks}, {Carnero Rosell}, {Carrasco Kind}, {Carretero}, {Costanzi}, {De Vicente}, {Desai}, {Diehl}, {Doel}, {Everett}, {Evrard}, {Ferrero}, {Fert{\'e}}, {Flaugher}, {Fosalba}, {Frieman}, {Garc{\'\i}a-Bellido}, {Giles}, {Gruendl}, {Gutierrez}, {Hinton}, {Hollowood}, {Honscheid}, {James}, {Jeltema}, {Kuehn}, {Kuropatkin}, {Lahav}, {Melchior}, {Miquel}, {Morgan}, {Palmese}, {Paz-Chinch{\'o}n}, {Plazas}, {Romer}, {Sanchez}, {Santiago}, {Schubnell}, {Serrano}, {Sevilla-Noarbe}, {Smith}, {Soares-Santos}, {Suchyta}, {Tarle}, {To}, {Tucker}, \& {Wilkinson}}]{aguena21}
{Aguena}, M., {Benoist}, C., {da Costa}, L.~N., {et~al.} 2021, \mnras, 502, 4435, \dodoi{10.1093/mnras/stab264}

\bibitem[{{Aihara} {et~al.}(2011){Aihara}, {Allende Prieto}, {An}, {Anderson}, {Aubourg}, {Balbinot}, {Beers}, {Berlind}, {Bickerton}, {Bizyaev}, {Blanton}, {Bochanski}, {Bolton}, {Bovy}, {Brandt}, {Brinkmann}, {Brown}, {Brownstein}, {Busca}, {Campbell}, {Carr}, {Chen}, {Chiappini}, {Comparat}, {Connolly}, {Cortes}, {Croft}, {Cuesta}, {da Costa}, {Davenport}, {Dawson}, {Dhital}, {Ealet}, {Ebelke}, {Edmondson}, {Eisenstein}, {Escoffier}, {Esposito}, {Evans}, {Fan}, {Femen{\'\i}a Castell{\'a}}, {Font-Ribera}, {Frinchaboy}, {Ge}, {Gillespie}, {Gilmore}, {Gonz{\'a}lez Hern{\'a}ndez}, {Gott}, {Gould}, {Grebel}, {Gunn}, {Hamilton}, {Harding}, {Harris}, {Hawley}, {Hearty}, {Ho}, {Hogg}, {Holtzman}, {Honscheid}, {Inada}, {Ivans}, {Jiang}, {Johnson}, {Jordan}, {Jordan}, {Kazin}, {Kirkby}, {Klaene}, {Knapp}, {Kneib}, {Kochanek}, {Koesterke}, {Kollmeier}, {Kron}, {Lampeitl}, {Lang}, {Le Goff}, {Lee}, {Lin}, {Long}, {Loomis}, {Lucatello}, {Lundgren}, {Lupton}, {Ma}, {MacDonald}, {Mahadevan}, {Maia}, {Makler},
  {Malanushenko}, {Malanushenko}, {Mandelbaum}, {Maraston}, {Margala}, {Masters}, {McBride}, {McGehee}, {McGreer}, {M{\'e}nard}, {Miralda-Escud{\'e}}, {Morrison}, {Mullally}, {Muna}, {Munn}, {Murayama}, {Myers}, {Naugle}, {Neto}, {Nguyen}, {Nichol}, {O'Connell}, {Ogando}, {Olmstead}, {Oravetz}, {Padmanabhan}, {Palanque-Delabrouille}, {Pan}, {Pandey}, {P{\^a}ris}, {Percival}, {Petitjean}, {Pfaffenberger}, {Pforr}, {Phleps}, {Pichon}, {Pieri}, {Prada}, {Price-Whelan}, {Raddick}, {Ramos}, {Reyl{\'e}}, {Rich}, {Richards}, {Rix}, {Robin}, {Rocha-Pinto}, {Rockosi}, {Roe}, {Rollinde}, {Ross}, {Ross}, {Rossetto}, {S{\'a}nchez}, {Sayres}, {Schlegel}, {Schlesinger}, {Schmidt}, {Schneider}, {Sheldon}, {Shu}, {Simmerer}, {Simmons}, {Sivarani}, {Snedden}, {Sobeck}, {Steinmetz}, {Strauss}, {Szalay}, {Tanaka}, {Thakar}, {Thomas}, {Tinker}, {Tofflemire}, {Tojeiro}, {Tremonti}, {Vandenberg}, {Vargas Maga{\~n}a}, {Verde}, {Vogt}, {Wake}, {Wang}, {Weaver}, {Weinberg}, {White}, {White}, {Yanny}, {Yasuda}, {Yeche}, \&
  {Zehavi}}]{aihara11}
{Aihara}, H., {Allende Prieto}, C., {An}, D., {et~al.} 2011, \apjs, 193, 29, \dodoi{10.1088/0067-0049/193/2/29}

\bibitem[{{Aihara} {et~al.}(2018){Aihara}, {Arimoto}, {Armstrong}, {Arnouts}, {Bahcall}, {Bickerton}, {Bosch}, {Bundy}, {Capak}, {Chan}, {Chiba}, {Coupon}, {Egami}, {Enoki}, {Finet}, {Fujimori}, {Fujimoto}, {Furusawa}, {Furusawa}, {Goto}, {Goulding}, {Greco}, {Greene}, {Gunn}, {Hamana}, {Harikane}, {Hashimoto}, {Hattori}, {Hayashi}, {Hayashi}, {He{\l}miniak}, {Higuchi}, {Hikage}, {Ho}, {Hsieh}, {Huang}, {Huang}, {Ikeda}, {Imanishi}, {Inoue}, {Iwasawa}, {Iwata}, {Jaelani}, {Jian}, {Kamata}, {Karoji}, {Kashikawa}, {Katayama}, {Kawanomoto}, {Kayo}, {Koda}, {Koike}, {Kojima}, {Komiyama}, {Konno}, {Koshida}, {Koyama}, {Kusakabe}, {Leauthaud}, {Lee}, {Lin}, {Lin}, {Lupton}, {Mandelbaum}, {Matsuoka}, {Medezinski}, {Mineo}, {Miyama}, {Miyatake}, {Miyazaki}, {Momose}, {More}, {More}, {Moritani}, {Moriya}, {Morokuma}, {Mukae}, {Murata}, {Murayama}, {Nagao}, {Nakata}, {Niida}, {Niikura}, {Nishizawa}, {Obuchi}, {Oguri}, {Oishi}, {Okabe}, {Okamoto}, {Okura}, {Ono}, {Onodera}, {Onoue}, {Osato}, {Ouchi}, {Price}, {Pyo},
  {Sako}, {Sawicki}, {Shibuya}, {Shimasaku}, {Shimono}, {Shirasaki}, {Silverman}, {Simet}, {Speagle}, {Spergel}, {Strauss}, {Sugahara}, {Sugiyama}, {Suto}, {Suyu}, {Suzuki}, {Tait}, {Takada}, {Takata}, {Tamura}, {Tanaka}, {Tanaka}, {Tanaka}, {Tanaka}, {Terai}, {Terashima}, {Toba}, {Tominaga}, {Toshikawa}, {Turner}, {Uchida}, {Uchiyama}, {Umetsu}, {Uraguchi}, {Urata}, {Usuda}, {Utsumi}, {Wang}, {Wang}, {Wong}, {Yabe}, {Yamada}, {Yamanoi}, {Yasuda}, {Yeh}, {Yonehara}, \& {Yuma}}]{aihara18}
{Aihara}, H., {Arimoto}, N., {Armstrong}, R., {et~al.} 2018, \pasj, 70, S4, \dodoi{10.1093/pasj/psx066}

\bibitem[{{Aihara} {et~al.}(2022){Aihara}, {AlSayyad}, {Ando}, {Armstrong}, {Bosch}, {Egami}, {Furusawa}, {Furusawa}, {Harasawa}, {Harikane}, {Hsieh}, {Ikeda}, {Ito}, {Iwata}, {Kodama}, {Koike}, {Kokubo}, {Komiyama}, {Li}, {Liang}, {Lin}, {Lupton}, {Lust}, {MacArthur}, {Mawatari}, {Mineo}, {Miyatake}, {Miyazaki}, {More}, {Morishima}, {Murayama}, {Nakajima}, {Nakata}, {Nishizawa}, {Oguri}, {Okabe}, {Okura}, {Ono}, {Osato}, {Ouchi}, {Pan}, {Plazas Malag{\'o}n}, {Price}, {Reed}, {Rykoff}, {Shibuya}, {Simunovic}, {Strauss}, {Sugimori}, {Suto}, {Suzuki}, {Takada}, {Takagi}, {Takata}, {Takita}, {Tanaka}, {Tang}, {Taranu}, {Terai}, {Toba}, {Turner}, {Uchiyama}, {Vijarnwannaluk}, {Waters}, {Yamada}, {Yamamoto}, \& {Yamashita}}]{aihara22}
{Aihara}, H., {AlSayyad}, Y., {Ando}, M., {et~al.} 2022, \pasj, 74, 247, \dodoi{10.1093/pasj/psab122}

\bibitem[{{Angulo} \& {White}(2010)}]{angulo10}
{Angulo}, R.~E., \& {White}, S.~D.~M. 2010, \mnras, 405, 143, \dodoi{10.1111/j.1365-2966.2010.16459.x}

\bibitem[{{Araya-Araya} {et~al.}(2021){Araya-Araya}, {Vicentin}, {Sodr{\'e}}, {Overzier}, \& {Cuevas}}]{araya21}
{Araya-Araya}, P., {Vicentin}, M.~C., {Sodr{\'e}}, Laerte, J., {Overzier}, R.~A., \& {Cuevas}, H. 2021, \mnras, 504, 5054, \dodoi{10.1093/mnras/stab1133}

\bibitem[{{Araya-Araya} {et~al.}(2024){Araya-Araya}, {Cochrane}, {Hayward}, {Yates}, {Sodr{\'e}}, {Vicentin}, {Rennehan}, {Overzier}, \& {van Daalen}}]{araya24}
{Araya-Araya}, P., {Cochrane}, R.~K., {Hayward}, C.~C., {et~al.} 2024, \apj, 977, 204, \dodoi{10.3847/1538-4357/ad90ae}

\bibitem[{{Astropy Collaboration} {et~al.}(2013){Astropy Collaboration}, {Robitaille}, {Tollerud}, {Greenfield}, {Droettboom}, {Bray}, {Aldcroft}, {Davis}, {Ginsburg}, {Price-Whelan}, {Kerzendorf}, {Conley}, {Crighton}, {Barbary}, {Muna}, {Ferguson}, {Grollier}, {Parikh}, {Nair}, {Unther}, {Deil}, {Woillez}, {Conseil}, {Kramer}, {Turner}, {Singer}, {Fox}, {Weaver}, {Zabalza}, {Edwards}, {Azalee Bostroem}, {Burke}, {Casey}, {Crawford}, {Dencheva}, {Ely}, {Jenness}, {Labrie}, {Lim}, {Pierfederici}, {Pontzen}, {Ptak}, {Refsdal}, {Servillat}, \& {Streicher}}]{astropy}
{Astropy Collaboration}, {Robitaille}, T.~P., {Tollerud}, E.~J., {et~al.} 2013, \aap, 558, A33, \dodoi{10.1051/0004-6361/201322068}

\bibitem[{{Bernardi}(2009)}]{bernardi09}
{Bernardi}, M. 2009, \mnras, 395, 1491, \dodoi{10.1111/j.1365-2966.2009.14601.x}

\bibitem[{{Bleem} {et~al.}(2015){Bleem}, {Stalder}, {de Haan}, {Aird}, {Allen}, {Applegate}, {Ashby}, {Bautz}, {Bayliss}, {Benson}, {Bocquet}, {Brodwin}, {Carlstrom}, {Chang}, {Chiu}, {Cho}, {Clocchiatti}, {Crawford}, {Crites}, {Desai}, {Dietrich}, {Dobbs}, {Foley}, {Forman}, {George}, {Gladders}, {Gonzalez}, {Halverson}, {Hennig}, {Hoekstra}, {Holder}, {Holzapfel}, {Hrubes}, {Jones}, {Keisler}, {Knox}, {Lee}, {Leitch}, {Liu}, {Lueker}, {Luong-Van}, {Mantz}, {Marrone}, {McDonald}, {McMahon}, {Meyer}, {Mocanu}, {Mohr}, {Murray}, {Padin}, {Pryke}, {Reichardt}, {Rest}, {Ruel}, {Ruhl}, {Saliwanchik}, {Saro}, {Sayre}, {Schaffer}, {Schrabback}, {Shirokoff}, {Song}, {Spieler}, {Stanford}, {Staniszewski}, {Stark}, {Story}, {Stubbs}, {Vanderlinde}, {Vieira}, {Vikhlinin}, {Williamson}, {Zahn}, \& {Zenteno}}]{bleem15}
{Bleem}, L.~E., {Stalder}, B., {de Haan}, T., {et~al.} 2015, \apjs, 216, 27, \dodoi{10.1088/0067-0049/216/2/27}

\bibitem[{{Bond} {et~al.}(1996){Bond}, {Kofman}, \& {Pogosyan}}]{bond96}
{Bond}, J.~R., {Kofman}, L., \& {Pogosyan}, D. 1996, \nat, 380, 603, \dodoi{10.1038/380603a0}

\bibitem[{{Bosch} {et~al.}(2018){Bosch}, {Armstrong}, {Bickerton}, {Furusawa}, {Ikeda}, {Koike}, {Lupton}, {Mineo}, {Price}, {Takata}, {Tanaka}, {Yasuda}, {AlSayyad}, {Becker}, {Coulton}, {Coupon}, {Garmilla}, {Huang}, {Krughoff}, {Lang}, {Leauthaud}, {Lim}, {Lust}, {MacArthur}, {Mandelbaum}, {Miyatake}, {Miyazaki}, {Murata}, {More}, {Okura}, {Owen}, {Swinbank}, {Strauss}, {Yamada}, \& {Yamanoi}}]{bosch18}
{Bosch}, J., {Armstrong}, R., {Bickerton}, S., {et~al.} 2018, \pasj, 70, S5, \dodoi{10.1093/pasj/psx080}

\bibitem[{{Bower} {et~al.}(1999){Bower}, {Terlevich}, {Kodama}, \& {Caldwell}}]{bower99}
{Bower}, R.~G., {Terlevich}, A., {Kodama}, T., \& {Caldwell}, N. 1999, in Astronomical Society of the Pacific Conference Series, Vol. 163, Star Formation in Early Type Galaxies, ed. P.~{Carral} \& J.~{Cepa}, 211, \dodoi{10.48550/arXiv.astro-ph/9808325}

\bibitem[{{Bruzual} \& {Charlot}(2003)}]{bruzual03}
{Bruzual}, G., \& {Charlot}, S. 2003, \mnras, 344, 1000, \dodoi{10.1046/j.1365-8711.2003.06897.x}

\bibitem[{{Chabrier}(2003)}]{chabrier03}
{Chabrier}, G. 2003, \pasp, 115, 763, \dodoi{10.1086/376392}

\bibitem[{{Chapman} {et~al.}(2000){Chapman}, {McCarthy}, \& {Persson}}]{chapman00}
{Chapman}, S.~C., {McCarthy}, P.~J., \& {Persson}, S.~E. 2000, \aj, 120, 1612, \dodoi{10.1086/301579}

\bibitem[{{Chiang} {et~al.}(2013){Chiang}, {Overzier}, \& {Gebhardt}}]{chiang13}
{Chiang}, Y.-K., {Overzier}, R., \& {Gebhardt}, K. 2013, \apj, 779, 127, \dodoi{10.1088/0004-637X/779/2/127}

\bibitem[{{Chiang} {et~al.}(2017){Chiang}, {Overzier}, {Gebhardt}, \& {Henriques}}]{chiang17}
{Chiang}, Y.-K., {Overzier}, R.~A., {Gebhardt}, K., \& {Henriques}, B. 2017, \apjl, 844, L23, \dodoi{10.3847/2041-8213/aa7e7b}

\bibitem[{{Chollet} \& {others}(2018)}]{chollet18}
{Chollet}, F., \& {others}. 2018, {Keras: The Python Deep Learning library}, Astrophysics Source Code Library, record ascl:1806.022

\bibitem[{{Contini}(2021)}]{contini21}
{Contini}, E. 2021, Galaxies, 9, 60, \dodoi{10.3390/galaxies9030060}

\bibitem[{{Crain} {et~al.}(2015){Crain}, {Schaye}, {Bower}, {Furlong}, {Schaller}, {Theuns}, {Dalla Vecchia}, {Frenk}, {McCarthy}, {Helly}, {Jenkins}, {Rosas-Guevara}, {White}, \& {Trayford}}]{crain15}
{Crain}, R.~A., {Schaye}, J., {Bower}, R.~G., {et~al.} 2015, \mnras, 450, 1937, \dodoi{10.1093/mnras/stv725}

\bibitem[{{Cucciati} {et~al.}(2018){Cucciati}, {Lemaux}, {Zamorani}, {Le F{\`e}vre}, {Tasca}, {Hathi}, {Lee}, {Bardelli}, {Cassata}, {Garilli}, {Le Brun}, {Maccagni}, {Pentericci}, {Thomas}, {Vanzella}, {Zucca}, {Lubin}, {Amorin}, {Cassar{\`a}}, {Cimatti}, {Talia}, {Vergani}, {Koekemoer}, {Pforr}, \& {Salvato}}]{cucciati18}
{Cucciati}, O., {Lemaux}, B.~C., {Zamorani}, G., {et~al.} 2018, \aap, 619, A49, \dodoi{10.1051/0004-6361/201833655}

\bibitem[{{Dalal} {et~al.}(2021){Dalal}, {Strauss}, {Sunayama}, {Oguri}, {Lin}, {Huang}, {Park}, \& {Takada}}]{dalal21}
{Dalal}, R., {Strauss}, M.~A., {Sunayama}, T., {et~al.} 2021, \mnras, 507, 4016, \dodoi{10.1093/mnras/stab2363}

\bibitem[{{Davis} {et~al.}(1985){Davis}, {Efstathiou}, {Frenk}, \& {White}}]{davis85}
{Davis}, M., {Efstathiou}, G., {Frenk}, C.~S., \& {White}, S.~D.~M. 1985, \apj, 292, 371, \dodoi{10.1086/163168}

\bibitem[{{Dawson} {et~al.}(2013){Dawson}, {Schlegel}, {Ahn}, {Anderson}, {Aubourg}, {Bailey}, {Barkhouser}, {Bautista}, {Beifiori}, {Berlind}, {Bhardwaj}, {Bizyaev}, {Blake}, {Blanton}, {Blomqvist}, {Bolton}, {Borde}, {Bovy}, {Brandt}, {Brewington}, {Brinkmann}, {Brown}, {Brownstein}, {Bundy}, {Busca}, {Carithers}, {Carnero}, {Carr}, {Chen}, {Comparat}, {Connolly}, {Cope}, {Croft}, {Cuesta}, {da Costa}, {Davenport}, {Delubac}, {de Putter}, {Dhital}, {Ealet}, {Ebelke}, {Eisenstein}, {Escoffier}, {Fan}, {Filiz Ak}, {Finley}, {Font-Ribera}, {G{\'e}nova-Santos}, {Gunn}, {Guo}, {Haggard}, {Hall}, {Hamilton}, {Harris}, {Harris}, {Ho}, {Hogg}, {Holder}, {Honscheid}, {Huehnerhoff}, {Jordan}, {Jordan}, {Kauffmann}, {Kazin}, {Kirkby}, {Klaene}, {Kneib}, {Le Goff}, {Lee}, {Long}, {Loomis}, {Lundgren}, {Lupton}, {Maia}, {Makler}, {Malanushenko}, {Malanushenko}, {Mandelbaum}, {Manera}, {Maraston}, {Margala}, {Masters}, {McBride}, {McDonald}, {McGreer}, {McMahon}, {Mena}, {Miralda-Escud{\'e}}, {Montero-Dorta},
  {Montesano}, {Muna}, {Myers}, {Naugle}, {Nichol}, {Noterdaeme}, {Nuza}, {Olmstead}, {Oravetz}, {Oravetz}, {Owen}, {Padmanabhan}, {Palanque-Delabrouille}, {Pan}, {Parejko}, {P{\^a}ris}, {Percival}, {P{\'e}rez-Fournon}, {P{\'e}rez-R{\`a}fols}, {Petitjean}, {Pfaffenberger}, {Pforr}, {Pieri}, {Prada}, {Price-Whelan}, {Raddick}, {Rebolo}, {Rich}, {Richards}, {Rockosi}, {Roe}, {Ross}, {Ross}, {Rossi}, {Rubi{\~n}o-Martin}, {Samushia}, {S{\'a}nchez}, {Sayres}, {Schmidt}, {Schneider}, {Sc{\'o}ccola}, {Seo}, {Shelden}, {Sheldon}, {Shen}, {Shu}, {Slosar}, {Smee}, {Snedden}, {Stauffer}, {Steele}, {Strauss}, {Streblyanska}, {Suzuki}, {Swanson}, {Tal}, {Tanaka}, {Thomas}, {Tinker}, {Tojeiro}, {Tremonti}, {Vargas Maga{\~n}a}, {Verde}, {Viel}, {Wake}, {Watson}, {Weaver}, {Weinberg}, {Weiner}, {West}, {White}, {Wood-Vasey}, {Yeche}, {Zehavi}, {Zhao}, \& {Zheng}}]{dawson13}
{Dawson}, K.~S., {Schlegel}, D.~J., {Ahn}, C.~P., {et~al.} 2013, \aj, 145, 10, \dodoi{10.1088/0004-6256/145/1/10}

\bibitem[{{De Lucia} \& {Blaizot}(2007)}]{delucia07}
{De Lucia}, G., \& {Blaizot}, J. 2007, \mnras, 375, 2, \dodoi{10.1111/j.1365-2966.2006.11287.x}

\bibitem[{{Doubrawa} {et~al.}(2024){Doubrawa}, {Cypriano}, {Finoguenov}, {Lopes}, {Gonzalez}, {Maturi}, {Dupke}, {Gonz{\'a}lez Delgado}, {Abramo}, {Benitez}, {Bonoli}, {Carneiro}, {Cenarro}, {Crist{\'o}bal-Hornillos}, {Ederoclite}, {Hern{\'a}n-Caballero}, {L{\'o}pez-Sanjuan}, {Mar{\'\i}n-Franch}, {Mendes de Oliveira}, {Moles}, {Sodr{\'e}}, {Taylor}, {Varela}, \& {V{\'a}zquez Rami{\'o}}}]{doubrawa24}
{Doubrawa}, L., {Cypriano}, E.~S., {Finoguenov}, A., {et~al.} 2024, \aap, 685, A98, \dodoi{10.1051/0004-6361/202349019}

\bibitem[{{Eisenstein} {et~al.}(2001){Eisenstein}, {Annis}, {Gunn}, {Szalay}, {Connolly}, {Nichol}, {Bahcall}, {Bernardi}, {Burles}, {Castander}, {Fukugita}, {Hogg}, {Ivezi{\'c}}, {Knapp}, {Lupton}, {Narayanan}, {Postman}, {Reichart}, {Richmond}, {Schneider}, {Schlegel}, {Strauss}, {SubbaRao}, {Tucker}, {Vanden Berk}, {Vogeley}, {Weinberg}, \& {Yanny}}]{eisenstein02}
{Eisenstein}, D.~J., {Annis}, J., {Gunn}, J.~E., {et~al.} 2001, \aj, 122, 2267, \dodoi{10.1086/323717}

\bibitem[{{Ferragamo} {et~al.}(2020){Ferragamo}, {Rubi{\~n}o-Mart{\'\i}n}, {Betancort-Rijo}, {Munari}, {Sartoris}, \& {Barrena}}]{ferragamo20}
{Ferragamo}, A., {Rubi{\~n}o-Mart{\'\i}n}, J.~A., {Betancort-Rijo}, J., {et~al.} 2020, \aap, 641, A41, \dodoi{10.1051/0004-6361/201834837}

\bibitem[{{Gobat} {et~al.}(2019){Gobat}, {Daddi}, {Coogan}, {Le Brun}, {Bournaud}, {Melin}, {Riechers}, {Sargent}, {Valentino}, {Hwang}, {Finoguenov}, \& {Strazzullo}}]{gobat19}
{Gobat}, R., {Daddi}, E., {Coogan}, R.~T., {et~al.} 2019, \aap, 629, A104, \dodoi{10.1051/0004-6361/201935862}

\bibitem[{{Gonzalez} {et~al.}(2019){Gonzalez}, {Gettings}, {Brodwin}, {Eisenhardt}, {Stanford}, {Wylezalek}, {Decker}, {Marrone}, {Moravec}, {O'Donnell}, {Stalder}, {Stern}, {Abdulla}, {Brown}, {Carlstrom}, {Chambers}, {Hayden}, {Lin}, {Magnier}, {Masci}, {Mantz}, {McDonald}, {Mo}, {Perlmutter}, {Wright}, \& {Zeimann}}]{gonzalez19}
{Gonzalez}, A.~H., {Gettings}, D.~P., {Brodwin}, M., {et~al.} 2019, \apjs, 240, 33, \dodoi{10.3847/1538-4365/aafad2}

\bibitem[{{Hao} {et~al.}(2010){Hao}, {McKay}, {Koester}, {Rykoff}, {Rozo}, {Annis}, {Wechsler}, {Evrard}, {Siegel}, {Becker}, {Busha}, {Gerdes}, {Johnston}, \& {Sheldon}}]{hao10}
{Hao}, J., {McKay}, T.~A., {Koester}, B.~P., {et~al.} 2010, \apjs, 191, 254, \dodoi{10.1088/0067-0049/191/2/254}

\bibitem[{{Harikane} {et~al.}(2018){Harikane}, {Ouchi}, {Ono}, {Saito}, {Behroozi}, {More}, {Shimasaku}, {Toshikawa}, {Lin}, {Akiyama}, {Coupon}, {Komiyama}, {Konno}, {Lin}, {Miyazaki}, {Nishizawa}, {Shibuya}, \& {Silverman}}]{harikane18}
{Harikane}, Y., {Ouchi}, M., {Ono}, Y., {et~al.} 2018, \pasj, 70, S11, \dodoi{10.1093/pasj/psx097}

\bibitem[{Harris {et~al.}(2020)Harris, Millman, van~der Walt, Gommers, Virtanen, Cournapeau, Wieser, Taylor, Berg, Smith, Kern, Picus, Hoyer, van Kerkwijk, Brett, Haldane, del R{\'{i}}o, Wiebe, Peterson, G{\'{e}}rard-Marchant, Sheppard, Reddy, Weckesser, Abbasi, Gohlke, \& Oliphant}]{harris2020}
Harris, C.~R., Millman, K.~J., van~der Walt, S.~J., {et~al.} 2020, Nature, 585, 357, \dodoi{10.1038/s41586-020-2649-2}

\bibitem[{{Henriques} {et~al.}(2015){Henriques}, {White}, {Thomas}, {Angulo}, {Guo}, {Lemson}, {Springel}, \& {Overzier}}]{henriques15}
{Henriques}, B. M.~B., {White}, S. D.~M., {Thomas}, P.~A., {et~al.} 2015, \mnras, 451, 2663, \dodoi{10.1093/mnras/stv705}

\bibitem[{{Hilton} {et~al.}(2021){Hilton}, {Sif{\'o}n}, {Naess}, {Madhavacheril}, {Oguri}, {Rozo}, {Rykoff}, {Abbott}, {Adhikari}, {Aguena}, {Aiola}, {Allam}, {Amodeo}, {Amon}, {Annis}, {Ansarinejad}, {Aros-Bunster}, {Austermann}, {Avila}, {Bacon}, {Battaglia}, {Beall}, {Becker}, {Bernstein}, {Bertin}, {Bhandarkar}, {Bhargava}, {Bond}, {Brooks}, {Burke}, {Calabrese}, {Carrasco Kind}, {Carretero}, {Choi}, {Choi}, {Conselice}, {da Costa}, {Costanzi}, {Crichton}, {Crowley}, {D{\"u}nner}, {Denison}, {Devlin}, {Dicker}, {Diehl}, {Dietrich}, {Doel}, {Duff}, {Duivenvoorden}, {Dunkley}, {Everett}, {Ferraro}, {Ferrero}, {Fert{\'e}}, {Flaugher}, {Frieman}, {Gallardo}, {Garc{\'\i}a-Bellido}, {Gaztanaga}, {Gerdes}, {Giles}, {Golec}, {Gralla}, {Grandis}, {Gruen}, {Gruendl}, {Gschwend}, {Gutierrez}, {Han}, {Hartley}, {Hasselfield}, {Hill}, {Hilton}, {Hincks}, {Hinton}, {Ho}, {Honscheid}, {Hoyle}, {Hubmayr}, {Huffenberger}, {Hughes}, {Jaelani}, {Jain}, {James}, {Jeltema}, {Kent}, {Knowles}, {Koopman}, {Kuehn}, {Lahav},
  {Lima}, {Lin}, {Lokken}, {Loubser}, {MacCrann}, {Maia}, {Marriage}, {Martin}, {McMahon}, {Melchior}, {Menanteau}, {Miquel}, {Miyatake}, {Moodley}, {Morgan}, {Mroczkowski}, {Nati}, {Newburgh}, {Niemack}, {Nishizawa}, {Ogando}, {Orlowski-Scherer}, {Page}, {Palmese}, {Partridge}, {Paz-Chinch{\'o}n}, {Phakathi}, {Plazas}, {Robertson}, {Romer}, {Carnero Rosell}, {Salatino}, {Sanchez}, {Schaan}, {Schillaci}, {Sehgal}, {Serrano}, {Shin}, {Simon}, {Smith}, {Soares-Santos}, {Spergel}, {Staggs}, {Storer}, {Suchyta}, {Swanson}, {Tarle}, {Thomas}, {To}, {Trac}, {Ullom}, {Vale}, {Van Lanen}, {Vavagiakis}, {De Vicente}, {Wilkinson}, {Wollack}, {Xu}, \& {Zhang}}]{hilton21}
{Hilton}, M., {Sif{\'o}n}, C., {Naess}, S., {et~al.} 2021, \apjs, 253, 3, \dodoi{10.3847/1538-4365/abd023}

\bibitem[{{Hoaglin} {et~al.}(1983){Hoaglin}, {Mosteller}, \& {Tukey}}]{hoaglin83}
{Hoaglin}, D.~C., {Mosteller}, F., \& {Tukey}, J.~W. 1983, {Understanding robust and exploratory data anlysis}

\bibitem[{{Huchra} \& {Geller}(1982)}]{huchra82}
{Huchra}, J.~P., \& {Geller}, M.~J. 1982, \apj, 257, 423, \dodoi{10.1086/160000}

\bibitem[{Hunter(2007)}]{hunter2007}
Hunter, J.~D. 2007, Computing in Science \& Engineering, 9, 90, \dodoi{10.1109/MCSE.2007.55}

\bibitem[{{Ivezi{\'c}} {et~al.}(2019){Ivezi{\'c}}, {Kahn}, {Tyson}, {Abel}, {Acosta}, {Allsman}, {Alonso}, {AlSayyad}, {Anderson}, {Andrew}, {Angel}, {Angeli}, {Ansari}, {Antilogus}, {Araujo}, {Armstrong}, {Arndt}, {Astier}, {Aubourg}, {Auza}, {Axelrod}, {Bard}, {Barr}, {Barrau}, {Bartlett}, {Bauer}, {Bauman}, {Baumont}, {Bechtol}, {Bechtol}, {Becker}, {Becla}, {Beldica}, {Bellavia}, {Bianco}, {Biswas}, {Blanc}, {Blazek}, {Blandford}, {Bloom}, {Bogart}, {Bond}, {Booth}, {Borgland}, {Borne}, {Bosch}, {Boutigny}, {Brackett}, {Bradshaw}, {Brandt}, {Brown}, {Bullock}, {Burchat}, {Burke}, {Cagnoli}, {Calabrese}, {Callahan}, {Callen}, {Carlin}, {Carlson}, {Chandrasekharan}, {Charles-Emerson}, {Chesley}, {Cheu}, {Chiang}, {Chiang}, {Chirino}, {Chow}, {Ciardi}, {Claver}, {Cohen-Tanugi}, {Cockrum}, {Coles}, {Connolly}, {Cook}, {Cooray}, {Covey}, {Cribbs}, {Cui}, {Cutri}, {Daly}, {Daniel}, {Daruich}, {Daubard}, {Daues}, {Dawson}, {Delgado}, {Dellapenna}, {de Peyster}, {de Val-Borro}, {Digel}, {Doherty}, {Dubois},
  {Dubois-Felsmann}, {Durech}, {Economou}, {Eifler}, {Eracleous}, {Emmons}, {Fausti Neto}, {Ferguson}, {Figueroa}, {Fisher-Levine}, {Focke}, {Foss}, {Frank}, {Freemon}, {Gangler}, {Gawiser}, {Geary}, {Gee}, {Geha}, {Gessner}, {Gibson}, {Gilmore}, {Glanzman}, {Glick}, {Goldina}, {Goldstein}, {Goodenow}, {Graham}, {Gressler}, {Gris}, {Guy}, {Guyonnet}, {Haller}, {Harris}, {Hascall}, {Haupt}, {Hernandez}, {Herrmann}, {Hileman}, {Hoblitt}, {Hodgson}, {Hogan}, {Howard}, {Huang}, {Huffer}, {Ingraham}, {Innes}, {Jacoby}, {Jain}, {Jammes}, {Jee}, {Jenness}, {Jernigan}, {Jevremovi{\'c}}, {Johns}, {Johnson}, {Johnson}, {Jones}, {Juramy-Gilles}, {Juri{\'c}}, {Kalirai}, {Kallivayalil}, {Kalmbach}, {Kantor}, {Karst}, {Kasliwal}, {Kelly}, {Kessler}, {Kinnison}, {Kirkby}, {Knox}, {Kotov}, {Krabbendam}, {Krughoff}, {Kub{\'a}nek}, {Kuczewski}, {Kulkarni}, {Ku}, {Kurita}, {Lage}, {Lambert}, {Lange}, {Langton}, {Le Guillou}, {Levine}, {Liang}, {Lim}, {Lintott}, {Long}, {Lopez}, {Lotz}, {Lupton}, {Lust}, {MacArthur}, {Mahabal},
  {Mandelbaum}, {Markiewicz}, {Marsh}, {Marshall}, {Marshall}, {May}, {McKercher}, {McQueen}, {Meyers}, {Migliore}, {Miller}, {Mills}, {Miraval}, {Moeyens}, {Moolekamp}, {Monet}, {Moniez}, {Monkewitz}, {Montgomery}, {Morrison}, {Mueller}, {Muller}, {Mu{\~n}oz Arancibia}, {Neill}, {Newbry}, {Nief}, {Nomerotski}, {Nordby}, {O'Connor}, {Oliver}, {Olivier}, {Olsen}, {O'Mullane}, {Ortiz}, {Osier}, {Owen}, {Pain}, {Palecek}, {Parejko}, {Parsons}, {Pease}, {Peterson}, {Peterson}, {Petravick}, {Libby Petrick}, {Petry}, {Pierfederici}, {Pietrowicz}, {Pike}, {Pinto}, {Plante}, {Plate}, {Plutchak}, {Price}, {Prouza}, {Radeka}, {Rajagopal}, {Rasmussen}, {Regnault}, {Reil}, {Reiss}, {Reuter}, {Ridgway}, {Riot}, {Ritz}, {Robinson}, {Roby}, {Roodman}, {Rosing}, {Roucelle}, {Rumore}, {Russo}, {Saha}, {Sassolas}, {Schalk}, {Schellart}, {Schindler}, {Schmidt}, {Schneider}, {Schneider}, {Schoening}, {Schumacher}, {Schwamb}, {Sebag}, {Selvy}, {Sembroski}, {Seppala}, {Serio}, {Serrano}, {Shaw}, {Shipsey}, {Sick}, {Silvestri},
  {Slater}, {Smith}, {Smith}, {Sobhani}, {Soldahl}, {Storrie-Lombardi}, {Stover}, {Strauss}, {Street}, {Stubbs}, {Sullivan}, {Sweeney}, {Swinbank}, {Szalay}, {Takacs}, {Tether}, {Thaler}, {Thayer}, {Thomas}, {Thornton}, {Thukral}, {Tice}, {Trilling}, {Turri}, {Van Berg}, {Vanden Berk}, {Vetter}, {Virieux}, {Vucina}, {Wahl}, {Walkowicz}, {Walsh}, {Walter}, {Wang}, {Wang}, {Warner}, {Wiecha}, {Willman}, {Winters}, {Wittman}, {Wolff}, {Wood-Vasey}, {Wu}, {Xin}, {Yoachim}, \& {Zhan}}]{ivezic19}
{Ivezi{\'c}}, {\v{Z}}., {Kahn}, S.~M., {Tyson}, J.~A., {et~al.} 2019, \apj, 873, 111, \dodoi{10.3847/1538-4357/ab042c}

\bibitem[{{Kawanomoto} {et~al.}(2018){Kawanomoto}, {Uraguchi}, {Komiyama}, {Miyazaki}, {Furusawa}, {Finet}, {Hattori}, {Wang}, {Yasuda}, \& {Suzuki}}]{kawanomoto18}
{Kawanomoto}, S., {Uraguchi}, F., {Komiyama}, Y., {et~al.} 2018, \pasj, 70, 66, \dodoi{10.1093/pasj/psy056}

\bibitem[{{Kitayama} {et~al.}(2023){Kitayama}, {Ueda}, {Okabe}, {Akahori}, {Hilton}, {Hughes}, {Ichinohe}, {Kohno}, {Komatsu}, {Lin}, {Miyatake}, {Oguri}, {Sif{\'o}n}, {Takakuwa}, {Takizawa}, {Tsutsumi}, {van Marrewijk}, \& {Wollack}}]{kitayama23}
{Kitayama}, T., {Ueda}, S., {Okabe}, N., {et~al.} 2023, \pasj, 75, 311, \dodoi{10.1093/pasj/psac110}

\bibitem[{{Klein} {et~al.}(2023){Klein}, {Hern{\'a}ndez-Lang}, {Mohr}, {Bocquet}, \& {Singh}}]{klein23}
{Klein}, M., {Hern{\'a}ndez-Lang}, D., {Mohr}, J.~J., {Bocquet}, S., \& {Singh}, A. 2023, \mnras, 526, 3757, \dodoi{10.1093/mnras/stad2729}

\bibitem[{{Koester} {et~al.}(2007){Koester}, {McKay}, {Annis}, {Wechsler}, {Evrard}, {Bleem}, {Becker}, {Johnston}, {Sheldon}, {Nichol}, {Miller}, {Scranton}, {Bahcall}, {Barentine}, {Brewington}, {Brinkmann}, {Harvanek}, {Kleinman}, {Krzesinski}, {Long}, {Nitta}, {Schneider}, {Sneddin}, {Voges}, \& {York}}]{koester07}
{Koester}, B.~P., {McKay}, T.~A., {Annis}, J., {et~al.} 2007, \apj, 660, 239, \dodoi{10.1086/509599}

\bibitem[{Koulouridis {et~al.}(2021)Koulouridis, Clerc, Sadibekova, Chira, Drigga, Faccioli, Le~F{\`e}vre, Garrel, Gaynullina, Gkini, {et~al.}}]{koulouridis21}
Koulouridis, E., Clerc, N., Sadibekova, T., {et~al.} 2021, Astronomy \& Astrophysics, 652, A12

\bibitem[{Kravtsov \& Borgani(2012)}]{kravtsov12}
Kravtsov, A.~V., \& Borgani, S. 2012, \araa, 50, 353, \dodoi{10.1146/annurev-astro-081811-125502}

\bibitem[{{Lauer} {et~al.}(2014){Lauer}, {Postman}, {Strauss}, {Graves}, \& {Chisari}}]{lauer14}
{Lauer}, T.~R., {Postman}, M., {Strauss}, M.~A., {Graves}, G.~J., \& {Chisari}, N.~E. 2014, \apj, 797, 82, \dodoi{10.1088/0004-637X/797/2/82}

\bibitem[{{Li} {et~al.}(2022){Li}, {Yang}, {Liu}, {Jing}, {He}, {Huang}, {Dai}, {Sawicki}, {Arnouts}, {Gwyn}, {Moutard}, {Mo}, {Wang}, {Katsianis}, {Cui}, {Han}, {Chiu}, {Gu}, \& {Xu}}]{qing22}
{Li}, Q., {Yang}, X., {Liu}, C., {et~al.} 2022, \apj, 933, 9, \dodoi{10.3847/1538-4357/ac6e69}

\bibitem[{{Lima} {et~al.}(2022){Lima}, {Sodr{\'e}}, {Bom}, {Teixeira}, {Nakazono}, {Buzzo}, {Queiroz}, {Herpich}, {Castellon}, {Dantas}, {Dors}, {Souza}, {Akras}, {Jim{\'e}nez-Teja}, {Kanaan}, {Ribeiro}, \& {Schoennell}}]{lima22}
{Lima}, E.~V.~R., {Sodr{\'e}}, L., {Bom}, C.~R., {et~al.} 2022, Astronomy and Computing, 38, 100510, \dodoi{10.1016/j.ascom.2021.100510}

\bibitem[{{Maraston}(2005)}]{maraston05}
{Maraston}, C. 2005, \mnras, 362, 799, \dodoi{10.1111/j.1365-2966.2005.09270.x}

\bibitem[{{Mendes de Oliveira} {et~al.}(2019){Mendes de Oliveira}, {Ribeiro}, {Schoenell}, {Kanaan}, {Overzier}, {Molino}, {Sampedro}, {Coelho}, {Barbosa}, {Cortesi}, {Costa-Duarte}, {Herpich}, {Hernandez-Jimenez}, {Placco}, {Xavier}, {Abramo}, {Saito}, {Chies-Santos}, {Ederoclite}, {Lopes de Oliveira}, {Gon{\c{c}}alves}, {Akras}, {Almeida}, {Almeida-Fernandes}, {Beers}, {Bonatto}, {Bonoli}, {Cypriano}, {Vinicius-Lima}, {de Souza}, {Fabiano de Souza}, {Ferrari}, {Gon{\c{c}}alves}, {Gonzalez}, {Guti{\'e}rrez-Soto}, {Hartmann}, {Jaffe}, {Kerber}, {Lima-Dias}, {Lopes}, {Menendez-Delmestre}, {Nakazono}, {Novais}, {Ortega-Minakata}, {Pereira}, {Perottoni}, {Queiroz}, {Reis}, {Santos}, {Santos-Silva}, {Santucci}, {Barbosa}, {Siffert}, {Sodr{\'e}}, {Torres-Flores}, {Westera}, {Whitten}, {Alcaniz}, {Alonso-Garc{\'\i}a}, {Alencar}, {Alvarez-Candal}, {Amram}, {Azanha}, {Barb{\'a}}, {Bernardinelli}, {Borges Fernandes}, {Branco}, {Brito-Silva}, {Buzzo}, {Caffer}, {Campillay}, {Cano}, {Carvano}, {Castejon}, {Cid
  Fernandes}, {Dantas}, {Daflon}, {Damke}, {de la Reza}, {de Melo de Azevedo}, {De Paula}, {Diem}, {Donnerstein}, {Dors}, {Dupke}, {Eikenberry}, {Escudero}, {Faifer}, {Far{\'\i}as}, {Fernandes}, {Fernandes}, {Fontes}, {Galarza}, {Hirata}, {Katena}, {Gregorio-Hetem}, {Hern{\'a}ndez-Fern{\'a}ndez}, {Izzo}, {Jaque Arancibia}, {Jatenco-Pereira}, {Jim{\'e}nez-Teja}, {Kann}, {Krabbe}, {Labayru}, {Lazzaro}, {Lima Neto}, {Lopes}, {Magalh{\~a}es}, {Makler}, {de Menezes}, {Miralda-Escud{\'e}}, {Monteiro-Oliveira}, {Montero-Dorta}, {Mu{\~n}oz-Elgueta}, {Nemmen}, {Nilo Castell{\'o}n}, {Oliveira}, {Ort{\'\i}z}, {Pattaro}, {Pereira}, {Quint}, {Riguccini}, {Rocha Pinto}, {Rodrigues}, {Roig}, {Rossi}, {Saha}, {Santos}, {Schnorr M{\"u}ller}, {Sesto}, {Silva}, {Smith Castelli}, {Teixeira}, {Telles}, {Thom de Souza}, {Th{\"o}ne}, {Trevisan}, {de Ugarte Postigo}, {Urrutia-Viscarra}, {Veiga}, {Vika}, {Vitorelli}, {Werle}, {Werner}, \& {Zaritsky}}]{mendesdeoliveira19}
{Mendes de Oliveira}, C., {Ribeiro}, T., {Schoenell}, W., {et~al.} 2019, \mnras, 489, 241, \dodoi{10.1093/mnras/stz1985}

\bibitem[{{Montenegro-Taborda} {et~al.}(2023){Montenegro-Taborda}, {Rodriguez-Gomez}, {Pillepich}, {Avila-Reese}, {Sales}, {Rodr{\'\i}guez-Puebla}, \& {Hernquist}}]{montenegro23}
{Montenegro-Taborda}, D., {Rodriguez-Gomez}, V., {Pillepich}, A., {et~al.} 2023, \mnras, 521, 800, \dodoi{10.1093/mnras/stad586}

\bibitem[{{Montes} \& {Trujillo}(2018)}]{montes18}
{Montes}, M., \& {Trujillo}, I. 2018, \mnras, 474, 917, \dodoi{10.1093/mnras/stx2847}

\bibitem[{{Nakata} {et~al.}(2001){Nakata}, {Kajisawa}, {Yamada}, {Kodama}, {Shimasaku}, {Tanaka}, {Doi}, {Furusawa}, {Hamabe}, {Iye}, {Kimura}, {Komiyama}, {Miyazaki}, {Okamura}, {Ouchi}, {Sasaki}, {Sekiguchi}, {Yagi}, \& {Yasuda}}]{nakata01}
{Nakata}, F., {Kajisawa}, M., {Yamada}, T., {et~al.} 2001, \pasj, 53, 1139, \dodoi{10.1093/pasj/53.6.1139}

\bibitem[{{Nishizawa} {et~al.}(2020){Nishizawa}, {Hsieh}, {Tanaka}, \& {Takata}}]{nishizawa20}
{Nishizawa}, A.~J., {Hsieh}, B.-C., {Tanaka}, M., \& {Takata}, T. 2020, arXiv e-prints, arXiv:2003.01511, \dodoi{10.48550/arXiv.2003.01511}

\bibitem[{{Oguri}(2014)}]{oguri14}
{Oguri}, M. 2014, \mnras, 444, 147, \dodoi{10.1093/mnras/stu1446}

\bibitem[{{Oguri} {et~al.}(2018){Oguri}, {Lin}, {Lin}, {Nishizawa}, {More}, {More}, {Hsieh}, {Medezinski}, {Miyatake}, {Jian}, {Lin}, {Takada}, {Okabe}, {Speagle}, {Coupon}, {Leauthaud}, {Lupton}, {Miyazaki}, {Price}, {Tanaka}, {Chiu}, {Komiyama}, {Okura}, {Tanaka}, \& {Usuda}}]{oguri18}
{Oguri}, M., {Lin}, Y.-T., {Lin}, S.-C., {et~al.} 2018, \pasj, 70, S20, \dodoi{10.1093/pasj/psx042}

\bibitem[{{Ota} {et~al.}(2023){Ota}, {Nguyen-Dang}, {Mitsuishi}, {Oguri}, {Klein}, {Okabe}, {Ramos-Ceja}, {Reiprich}, {Pacaud}, {Bulbul}, {Br{\"u}ggen}, {Liu}, {Migkas}, {Chiu}, {Ghirardini}, {Grandis}, {Lin}, {Miyatake}, {Miyazaki}, \& {Sanders}}]{ota23}
{Ota}, N., {Nguyen-Dang}, N.~T., {Mitsuishi}, I., {et~al.} 2023, \aap, 669, A110, \dodoi{10.1051/0004-6361/202244260}

\bibitem[{{Overzier}(2016)}]{overzier16}
{Overzier}, R.~A. 2016, \aapr, 24, 14, \dodoi{10.1007/s00159-016-0100-3}

\bibitem[{pandas~development team(2020)}]{reback2020}
pandas~development team, T. 2020, pandas-dev/pandas: Pandas, latest,  Zenodo, \dodoi{10.5281/zenodo.3509134}

\bibitem[{Pedregosa {et~al.}(2011)Pedregosa, Varoquaux, Gramfort, Michel, Thirion, Grisel, Blondel, Prettenhofer, Weiss, Dubourg, {et~al.}}]{pedregosa2011}
Pedregosa, F., Varoquaux, G., Gramfort, A., {et~al.} 2011, Journal of machine learning research, 12, 2825

\bibitem[{{Peebles}(1980)}]{peebles80}
{Peebles}, P.~J.~E. 1980, {The large-scale structure of the universe}

\bibitem[{Piffaretti {et~al.}(2011)Piffaretti, Arnaud, Pratt, Pointecouteau, \& Melin}]{piffaretti11}
Piffaretti, R., Arnaud, M., Pratt, G., Pointecouteau, E., \& Melin, J.-B. 2011, Astronomy \& Astrophysics, 534, A109

\bibitem[{{Planck Collaboration} {et~al.}(2014){Planck Collaboration}, {Ade}, {Aghanim}, {Armitage-Caplan}, {Arnaud}, {Ashdown}, {Atrio-Barand ela}, {Aumont}, {Baccigalupi}, {Banday}, {Barreiro}, {Bartlett}, {Battaner}, {Benabed}, {Beno{\^\i}t}, {Benoit-L{\'e}vy}, {Bernard}, {Bersanelli}, {Bielewicz}, {Bobin}, {Bock}, {Bonaldi}, {Bond}, {Borrill}, {Bouchet}, {Bridges}, {Bucher}, {Burigana}, {Butler}, {Calabrese}, {Cappellini}, {Cardoso}, {Catalano}, {Challinor}, {Chamballu}, {Chary}, {Chen}, {Chiang}, {Chiang}, {Christensen}, {Church}, {Clements}, {Colombi}, {Colombo}, {Couchot}, {Coulais}, {Crill}, {Curto}, {Cuttaia}, {Danese}, {Davies}, {Davis}, {de Bernardis}, {de Rosa}, {de Zotti}, {Delabrouille}, {Delouis}, {D{\'e}sert}, {Dickinson}, {Diego}, {Dolag}, {Dole}, {Donzelli}, {Dor{\'e}}, {Douspis}, {Dunkley}, {Dupac}, {Efstathiou}, {Elsner}, {En{\ss}lin}, {Eriksen}, {Finelli}, {Forni}, {Frailis}, {Fraisse}, {Franceschi}, {Gaier}, {Galeotta}, {Galli}, {Ganga}, {Giard}, {Giardino}, {Giraud-H{\'e}raud},
  {Gjerl{\o}w}, {Gonz{\'a}lez-Nuevo}, {G{\'o}rski}, {Gratton}, {Gregorio}, {Gruppuso}, {Gudmundsson}, {Haissinski}, {Hamann}, {Hansen}, {Hanson}, {Harrison}, {Henrot-Versill{\'e}}, {Hern{\'a}ndez-Monteagudo}, {Herranz}, {Hildebrand t}, {Hivon}, {Hobson}, {Holmes}, {Hornstrup}, {Hou}, {Hovest}, {Huffenberger}, {Jaffe}, {Jaffe}, {Jewell}, {Jones}, {Juvela}, {Keih{\"a}nen}, {Keskitalo}, {Kisner}, {Kneissl}, {Knoche}, {Knox}, {Kunz}, {Kurki-Suonio}, {Lagache}, {L{\"a}hteenm{\"a}ki}, {Lamarre}, {Lasenby}, {Lattanzi}, {Laureijs}, {Lawrence}, {Leach}, {Leahy}, {Leonardi}, {Le{\'o}n-Tavares}, {Lesgourgues}, {Lewis}, {Liguori}, {Lilje}, {Linden-V{\o}rnle}, {L{\'o}pez-Caniego}, {Lubin}, {Mac{\'\i}as-P{\'e}rez}, {Maffei}, {Maino}, {Mand olesi}, {Maris}, {Marshall}, {Martin}, {Mart{\'\i}nez-Gonz{\'a}lez}, {Masi}, {Massardi}, {Matarrese}, {Matthai}, {Mazzotta}, {Meinhold}, {Melchiorri}, {Melin}, {Mendes}, {Menegoni}, {Mennella}, {Migliaccio}, {Millea}, {Mitra}, {Miville-Desch{\^e}nes}, {Moneti}, {Montier}, {Morgante},
  {Mortlock}, {Moss}, {Munshi}, {Murphy}, {Naselsky}, {Nati}, {Natoli}, {Netterfield}, {N{\o}rgaard-Nielsen}, {Noviello}, {Novikov}, {Novikov}, {O'Dwyer}, {Osborne}, {Oxborrow}, {Paci}, {Pagano}, {Pajot}, {Paladini}, {Paoletti}, {Partridge}, {Pasian}, {Patanchon}, {Pearson}, {Pearson}, {Peiris}, {Perdereau}, {Perotto}, {Perrotta}, {Pettorino}, {Piacentini}, {Piat}, {Pierpaoli}, {Pietrobon}, {Plaszczynski}, {Platania}, {Pointecouteau}, {Polenta}, {Ponthieu}, {Popa}, {Poutanen}, {Pratt}, {Pr{\'e}zeau}, {Prunet}, {Puget}, {Rachen}, {Reach}, {Rebolo}, {Reinecke}, {Remazeilles}, {Renault}, {Ricciardi}, {Riller}, {Ristorcelli}, {Rocha}, {Rosset}, {Roudier}, {Rowan-Robinson}, {Rubi{\~n}o-Mart{\'\i}n}, {Rusholme}, {Sandri}, {Santos}, {Savelainen}, {Savini}, {Scott}, {Seiffert}, {Shellard}, {Spencer}, {Starck}, {Stolyarov}, {Stompor}, {Sudiwala}, {Sunyaev}, {Sureau}, {Sutton}, {Suur-Uski}, {Sygnet}, {Tauber}, {Tavagnacco}, {Terenzi}, {Toffolatti}, {Tomasi}, {Tristram}, {Tucci}, {Tuovinen}, {T{\"u}rler}, {Umana},
  {Valenziano}, {Valiviita}, {Van Tent}, {Vielva}, {Villa}, {Vittorio}, {Wade}, {Wandelt}, {Wehus}, {White}, {White}, {Wilkinson}, {Yvon}, {Zacchei}, \& {Zonca}}]{planck1}
{Planck Collaboration}, {Ade}, P.~A.~R., {Aghanim}, N., {et~al.} 2014, \aap, 571, A16, \dodoi{10.1051/0004-6361/201321591}

\bibitem[{{Planck Collaboration} {et~al.}(2016){Planck Collaboration}, {Ade}, {Aghanim}, {Arnaud}, {Ashdown}, {Aumont}, {Baccigalupi}, {Banday}, {Barreiro}, {Barrena}, {Bartlett}, {Bartolo}, {Battaner}, {Battye}, {Benabed}, {Beno{\^\i}t}, {Benoit-L{\'e}vy}, {Bernard}, {Bersanelli}, {Bielewicz}, {Bikmaev}, {B{\"o}hringer}, {Bonaldi}, {Bonavera}, {Bond}, {Borrill}, {Bouchet}, {Bucher}, {Burenin}, {Burigana}, {Butler}, {Calabrese}, {Cardoso}, {Carvalho}, {Catalano}, {Challinor}, {Chamballu}, {Chary}, {Chiang}, {Chon}, {Christensen}, {Clements}, {Colombi}, {Colombo}, {Combet}, {Comis}, {Couchot}, {Coulais}, {Crill}, {Curto}, {Cuttaia}, {Dahle}, {Danese}, {Davies}, {Davis}, {de Bernardis}, {de Rosa}, {de Zotti}, {Delabrouille}, {D{\'e}sert}, {Dickinson}, {Diego}, {Dolag}, {Dole}, {Donzelli}, {Dor{\'e}}, {Douspis}, {Ducout}, {Dupac}, {Efstathiou}, {Eisenhardt}, {Elsner}, {En{\ss}lin}, {Eriksen}, {Falgarone}, {Fergusson}, {Feroz}, {Ferragamo}, {Finelli}, {Forni}, {Frailis}, {Fraisse}, {Franceschi}, {Frejsel},
  {Galeotta}, {Galli}, {Ganga}, {G{\'e}nova-Santos}, {Giard}, {Giraud-H{\'e}raud}, {Gjerl{\o}w}, {Gonz{\'a}lez-Nuevo}, {G{\'o}rski}, {Grainge}, {Gratton}, {Gregorio}, {Gruppuso}, {Gudmundsson}, {Hansen}, {Hanson}, {Harrison}, {Hempel}, {Henrot-Versill{\'e}}, {Hern{\'a}ndez-Monteagudo}, {Herranz}, {Hildebrandt}, {Hivon}, {Hobson}, {Holmes}, {Hornstrup}, {Hovest}, {Huffenberger}, {Hurier}, {Jaffe}, {Jaffe}, {Jin}, {Jones}, {Juvela}, {Keih{\"a}nen}, {Keskitalo}, {Khamitov}, {Kisner}, {Kneissl}, {Knoche}, {Kunz}, {Kurki-Suonio}, {Lagache}, {Lamarre}, {Lasenby}, {Lattanzi}, {Lawrence}, {Leonardi}, {Lesgourgues}, {Levrier}, {Liguori}, {Lilje}, {Linden-V{\o}rnle}, {L{\'o}pez-Caniego}, {Lubin}, {Mac{\'\i}as-P{\'e}rez}, {Maggio}, {Maino}, {Mak}, {Mandolesi}, {Mangilli}, {Martin}, {Mart{\'\i}nez-Gonz{\'a}lez}, {Masi}, {Matarrese}, {Mazzotta}, {McGehee}, {Mei}, {Melchiorri}, {Melin}, {Mendes}, {Mennella}, {Migliaccio}, {Mitra}, {Miville-Desch{\^e}nes}, {Moneti}, {Montier}, {Morgante}, {Mortlock}, {Moss}, {Munshi},
  {Murphy}, {Naselsky}, {Nastasi}, {Nati}, {Natoli}, {Netterfield}, {N{\o}rgaard-Nielsen}, {Noviello}, {Novikov}, {Novikov}, {Olamaie}, {Oxborrow}, {Paci}, {Pagano}, {Pajot}, {Paoletti}, {Pasian}, {Patanchon}, {Pearson}, {Perdereau}, {Perotto}, {Perrott}, {Perrotta}, {Pettorino}, {Piacentini}, {Piat}, {Pierpaoli}, {Pietrobon}, {Plaszczynski}, {Pointecouteau}, {Polenta}, {Pratt}, {Pr{\'e}zeau}, {Prunet}, {Puget}, {Rachen}, {Reach}, {Rebolo}, {Reinecke}, {Remazeilles}, {Renault}, {Renzi}, {Ristorcelli}, {Rocha}, {Rosset}, {Rossetti}, {Roudier}, {Rozo}, {Rubi{\~n}o-Mart{\'\i}n}, {Rumsey}, {Rusholme}, {Rykoff}, {Sandri}, {Santos}, {Saunders}, {Savelainen}, {Savini}, {Schammel}, {Scott}, {Seiffert}, {Shellard}, {Shimwell}, {Spencer}, {Stanford}, {Stern}, {Stolyarov}, {Stompor}, {Streblyanska}, {Sudiwala}, {Sunyaev}, {Sutton}, {Suur-Uski}, {Sygnet}, {Tauber}, {Terenzi}, {Toffolatti}, {Tomasi}, {Tramonte}, {Tristram}, {Tucci}, {Tuovinen}, {Umana}, {Valenziano}, {Valiviita}, {Van Tent}, {Vielva}, {Villa}, {Wade},
  {Wandelt}, {Wehus}, {White}, {Wright}, {Yvon}, {Zacchei}, \& {Zonca}}]{ade16}
---. 2016, \aap, 594, A27, \dodoi{10.1051/0004-6361/201525823}

\bibitem[{{Postman} \& {Lauer}(1995)}]{postman95}
{Postman}, M., \& {Lauer}, T.~R. 1995, \apj, 440, 28, \dodoi{10.1086/175245}

\bibitem[{{Rykoff} {et~al.}(2014){Rykoff}, {Rozo}, {Busha}, {Cunha}, {Finoguenov}, {Evrard}, {Hao}, {Koester}, {Leauthaud}, {Nord}, {Pierre}, {Reddick}, {Sadibekova}, {Sheldon}, \& {Wechsler}}]{rykoff14}
{Rykoff}, E.~S., {Rozo}, E., {Busha}, M.~T., {et~al.} 2014, \apj, 785, 104, \dodoi{10.1088/0004-637X/785/2/104}

\bibitem[{{Rykoff} {et~al.}(2016){Rykoff}, {Rozo}, {Hollowood}, {Bermeo-Hernandez}, {Jeltema}, {Mayers}, {Romer}, {Rooney}, {Saro}, {Vergara Cervantes}, {Wechsler}, {Wilcox}, {Abbott}, {Abdalla}, {Allam}, {Annis}, {Benoit-L{\'e}vy}, {Bernstein}, {Bertin}, {Brooks}, {Burke}, {Capozzi}, {Carnero Rosell}, {Carrasco Kind}, {Castander}, {Childress}, {Collins}, {Cunha}, {D'Andrea}, {da Costa}, {Davis}, {Desai}, {Diehl}, {Dietrich}, {Doel}, {Evrard}, {Finley}, {Flaugher}, {Fosalba}, {Frieman}, {Glazebrook}, {Goldstein}, {Gruen}, {Gruendl}, {Gutierrez}, {Hilton}, {Honscheid}, {Hoyle}, {James}, {Kay}, {Kuehn}, {Kuropatkin}, {Lahav}, {Lewis}, {Lidman}, {Lima}, {Maia}, {Mann}, {Marshall}, {Martini}, {Melchior}, {Miller}, {Miquel}, {Mohr}, {Nichol}, {Nord}, {Ogando}, {Plazas}, {Reil}, {Sahl{\'e}n}, {Sanchez}, {Santiago}, {Scarpine}, {Schubnell}, {Sevilla-Noarbe}, {Smith}, {Soares-Santos}, {Sobreira}, {Stott}, {Suchyta}, {Swanson}, {Tarle}, {Thomas}, {Tucker}, {Uddin}, {Viana}, {Vikram}, {Walker}, {Zhang}, \& {DES
  Collaboration}}]{rykoff16}
{Rykoff}, E.~S., {Rozo}, E., {Hollowood}, D., {et~al.} 2016, \apjs, 224, 1, \dodoi{10.3847/0067-0049/224/1/1}

\bibitem[{Sarazin(1986)}]{sarazin86}
Sarazin, C.~L. 1986, Reviews of Modern Physics, 58, 1

\bibitem[{{Schlafly} {et~al.}(2018){Schlafly}, {Green}, {Lang}, {Daylan}, {Finkbeiner}, {Lee}, {Meisner}, {Schlegel}, \& {Valdes}}]{schlafly18}
{Schlafly}, E.~F., {Green}, G.~M., {Lang}, D., {et~al.} 2018, \apjs, 234, 39, \dodoi{10.3847/1538-4365/aaa3e2}

\bibitem[{{Shamshiri} {et~al.}(2015){Shamshiri}, {Thomas}, {Henriques}, {Tojeiro}, {Lemson}, {Oliver}, \& {Wilkins}}]{shamshiri15}
{Shamshiri}, S., {Thomas}, P.~A., {Henriques}, B.~M., {et~al.} 2015, \mnras, 451, 2681, \dodoi{10.1093/mnras/stv883}

\bibitem[{{Shimakawa} {et~al.}(2018){Shimakawa}, {Kodama}, {Hayashi}, {Prochaska}, {Tanaka}, {Cai}, {Suzuki}, {Tadaki}, \& {Koyama}}]{shimakawa18}
{Shimakawa}, R., {Kodama}, T., {Hayashi}, M., {et~al.} 2018, \mnras, 473, 1977, \dodoi{10.1093/mnras/stx2494}

\bibitem[{{Springel}(2005)}]{springel05}
{Springel}, V. 2005, \mnras, 364, 1105, \dodoi{10.1111/j.1365-2966.2005.09655.x}

\bibitem[{{Springel} {et~al.}(2001){Springel}, {White}, {Tormen}, \& {Kauffmann}}]{springel01}
{Springel}, V., {White}, S. D.~M., {Tormen}, G., \& {Kauffmann}, G. 2001, \mnras, 328, 726, \dodoi{10.1046/j.1365-8711.2001.04912.x}

\bibitem[{{Staniszewski} {et~al.}(2009){Staniszewski}, {Ade}, {Aird}, {Benson}, {Bleem}, {Carlstrom}, {Chang}, {Cho}, {Crawford}, {Crites}, {de Haan}, {Dobbs}, {Halverson}, {Holder}, {Holzapfel}, {Hrubes}, {Joy}, {Keisler}, {Lanting}, {Lee}, {Leitch}, {Loehr}, {Lueker}, {McMahon}, {Mehl}, {Meyer}, {Mohr}, {Montroy}, {Ngeow}, {Padin}, {Plagge}, {Pryke}, {Reichardt}, {Ruhl}, {Schaffer}, {Shaw}, {Shirokoff}, {Spieler}, {Stalder}, {Stark}, {Vanderlinde}, {Vieira}, {Zahn}, \& {Zenteno}}]{stanis09}
{Staniszewski}, Z., {Ade}, P.~A.~R., {Aird}, K.~A., {et~al.} 2009, \apj, 701, 32, \dodoi{10.1088/0004-637X/701/1/32}

\bibitem[{{Steidel} {et~al.}(2000){Steidel}, {Adelberger}, {Shapley}, {Pettini}, {Dickinson}, \& {Giavalisco}}]{steidel00}
{Steidel}, C.~C., {Adelberger}, K.~L., {Shapley}, A.~E., {et~al.} 2000, \apj, 532, 170, \dodoi{10.1086/308568}

\bibitem[{{Sunyaev} \& {Zeldovich}(1970)}]{sunyaev70}
{Sunyaev}, R.~A., \& {Zeldovich}, Y.~B. 1970, \apss, 7, 3, \dodoi{10.1007/BF00653471}

\bibitem[{{Takada} {et~al.}(2014){Takada}, {Ellis}, {Chiba}, {Greene}, {Aihara}, {Arimoto}, {Bundy}, {Cohen}, {Dor{\'e}}, {Graves}, {Gunn}, {Heckman}, {Hirata}, {Ho}, {Kneib}, {Le F{\`e}vre}, {Lin}, {More}, {Murayama}, {Nagao}, {Ouchi}, {Seiffert}, {Silverman}, {Sodr{\'e}}, {Spergel}, {Strauss}, {Sugai}, {Suto}, {Takami}, \& {Wyse}}]{takada14}
{Takada}, M., {Ellis}, R.~S., {Chiba}, M., {et~al.} 2014, \pasj, 66, R1, \dodoi{10.1093/pasj/pst019}

\bibitem[{{Tamura} {et~al.}(2016){Tamura}, {Takato}, {Shimono}, {Moritani}, {Yabe}, {Ishizuka}, {Ueda}, {Kamata}, {Aghazarian}, {Arnouts}, {Barban}, {Barkhouser}, {Borges}, {Braun}, {Carr}, {Chabaud}, {Chang}, {Chen}, {Chiba}, {Chou}, {Chu}, {Cohen}, {de Almeida}, {de Oliveira}, {de Oliveira}, {Dekany}, {Dohlen}, {dos Santos}, {dos Santos}, {Ellis}, {Fabricius}, {Ferrand}, {Ferreira}, {Golebiowski}, {Greene}, {Gross}, {Gunn}, {Hammond}, {Harding}, {Hart}, {Heckman}, {Hirata}, {Ho}, {Hope}, {Hovland}, {Hsu}, {Hu}, {Huang}, {Jaquet}, {Jing}, {Karr}, {Kimura}, {King}, {Komatsu}, {Le Brun}, {Le F{\`e}vre}, {Le Fur}, {Le Mignant}, {Ling}, {Loomis}, {Lupton}, {Madec}, {Mao}, {Marrara}, {Mendes de Oliveira}, {Minowa}, {Morantz}, {Murayama}, {Murray}, {Ohyama}, {Orndorff}, {Pascal}, {Pereira}, {Reiley}, {Reinecke}, {Ritter}, {Roberts}, {Schwochert}, {Seiffert}, {Smee}, {Sodre}, {Spergel}, {Steinkraus}, {Strauss}, {Surace}, {Suto}, {Suzuki}, {Swinbank}, {Tait}, {Takada}, {Tamura}, {Tanaka}, {Tresse}, {Verducci},
  {Vibert}, {Vidal}, {Wang}, {Wen}, {Yan}, \& {Yasuda}}]{tamura16}
{Tamura}, N., {Takato}, N., {Shimono}, A., {et~al.} 2016, in Society of Photo-Optical Instrumentation Engineers (SPIE) Conference Series, Vol. 9908, Ground-based and Airborne Instrumentation for Astronomy VI, ed. C.~J. {Evans}, L.~{Simard}, \& H.~{Takami}, 99081M, \dodoi{10.1117/12.2232103}

\bibitem[{{Tanaka} {et~al.}(2018){Tanaka}, {Coupon}, {Hsieh}, {Mineo}, {Nishizawa}, {Speagle}, {Furusawa}, {Miyazaki}, \& {Murayama}}]{tanaka18}
{Tanaka}, M., {Coupon}, J., {Hsieh}, B.-C., {et~al.} 2018, \pasj, 70, S9, \dodoi{10.1093/pasj/psx077}

\bibitem[{{The Dark Energy Survey Collaboration}(2005)}]{des05}
{The Dark Energy Survey Collaboration}. 2005, arXiv e-prints, astro, \dodoi{10.48550/arXiv.astro-ph/0510346}

\bibitem[{{Toshikawa} {et~al.}(2018){Toshikawa}, {Uchiyama}, {Kashikawa}, {Ouchi}, {Overzier}, {Ono}, {Harikane}, {Ishikawa}, {Kodama}, {Matsuda}, {Lin}, {Onoue}, {Tanaka}, {Nagao}, {Akiyama}, {Komiyama}, {Goto}, \& {Lee}}]{toshikawa18}
{Toshikawa}, J., {Uchiyama}, H., {Kashikawa}, N., {et~al.} 2018, \pasj, 70, S12, \dodoi{10.1093/pasj/psx102}

\bibitem[{{van Dokkum} {et~al.}(2010){van Dokkum}, {Whitaker}, {Brammer}, {Franx}, {Kriek}, {Labb{\'e}}, {Marchesini}, {Quadri}, {Bezanson}, {Illingworth}, {Muzzin}, {Rudnick}, {Tal}, \& {Wake}}]{vandokkum10}
{van Dokkum}, P.~G., {Whitaker}, K.~E., {Brammer}, G., {et~al.} 2010, \apj, 709, 1018, \dodoi{10.1088/0004-637X/709/2/1018}

\bibitem[{{Vicentin} {et~al.}(2025){Vicentin}, {Sodr{\'e}}, {Strauss}, {de Lima}, \& {Araya-Araya}}]{vicentinII25}
{Vicentin}, M.~C., {Sodr{\'e}}, Jr., L., {Strauss}, M.~A., {de Lima}, E. V.~R., \& {Araya-Araya}, P. 2025, \apj, 992, 53, \dodoi{10.3847/1538-4357/adff72}

\bibitem[{Virtanen {et~al.}(2020)Virtanen, Gommers, Oliphant, Haberland, Reddy, Cournapeau, Burovski, Peterson, Weckesser, Bright, {van der Walt}, Brett, Wilson, Millman, Mayorov, Nelson, Jones, Kern, Larson, Carey, Polat, Feng, Moore, {VanderPlas}, Laxalde, Perktold, Cimrman, Henriksen, Quintero, Harris, Archibald, Ribeiro, Pedregosa, {van Mulbregt}, \& {SciPy 1.0 Contributors}}]{scipy}
Virtanen, P., Gommers, R., Oliphant, T.~E., {et~al.} 2020, Nature Methods, 17, 261, \dodoi{10.1038/s41592-019-0686-2}

\bibitem[{{Weaver} {et~al.}(2022){Weaver}, {Kauffmann}, {Ilbert}, {McCracken}, {Moneti}, {Toft}, {Brammer}, {Shuntov}, {Davidzon}, {Hsieh}, {Laigle}, {Anastasiou}, {Jespersen}, {Vinther}, {Capak}, {Casey}, {McPartland}, {Milvang-Jensen}, {Mobasher}, {Sanders}, {Zalesky}, {Arnouts}, {Aussel}, {Dunlop}, {Faisst}, {Franx}, {Furtak}, {Fynbo}, {Gould}, {Greve}, {Gwyn}, {Kartaltepe}, {Kashino}, {Koekemoer}, {Kokorev}, {Le F{\`e}vre}, {Lilly}, {Masters}, {Magdis}, {Mehta}, {Peng}, {Riechers}, {Salvato}, {Sawicki}, {Scarlata}, {Scoville}, {Shirley}, {Silverman}, {Sneppen}, {Smolc̆i{\'c}}, {Steinhardt}, {Stern}, {Tanaka}, {Taniguchi}, {Teplitz}, {Vaccari}, {Wang}, \& {Zamorani}}]{weaver22}
{Weaver}, J.~R., {Kauffmann}, O.~B., {Ilbert}, O., {et~al.} 2022, \apjs, 258, 11, \dodoi{10.3847/1538-4365/ac3078}

\bibitem[{{Wen} \& {Han}(2015)}]{wh15}
{Wen}, Z.~L., \& {Han}, J.~L. 2015, \apj, 807, 178, \dodoi{10.1088/0004-637X/807/2/178}

\bibitem[{{Wen} \& {Han}(2021)}]{wh21}
---. 2021, \mnras, 500, 1003, \dodoi{10.1093/mnras/staa3308}

\bibitem[{{Wen} {et~al.}(2012){Wen}, {Han}, \& {Liu}}]{wh12}
{Wen}, Z.~L., {Han}, J.~L., \& {Liu}, F.~S. 2012, \apjs, 199, 34, \dodoi{10.1088/0067-0049/199/2/34}

\bibitem[{{Werner} {et~al.}(2022){Werner}, {Hatch}, {Muzzin}, {van der Burg}, {Balogh}, {Rudnick}, \& {Wilson}}]{werner22}
{Werner}, S.~V., {Hatch}, N.~A., {Muzzin}, A., {et~al.} 2022, \mnras, 510, 674, \dodoi{10.1093/mnras/stab3484}

\bibitem[{{Werner} {et~al.}(2023){Werner}, {Cypriano}, {Gonzalez}, {Mendes de Oliveira}, {Araya-Araya}, {Doubrawa}, {Lopes de Oliveira}, {Lopes}, {Vitorelli}, {Brambila}, {Costa-Duarte}, {Telles}, {Kanaan}, {Ribeiro}, {Schoenell}, {Gon{\c{c}}alves}, {Men{\'e}ndez-Delmestre}, {Bom}, \& {Nakazono}}]{werner23}
{Werner}, S.~V., {Cypriano}, E.~S., {Gonzalez}, A.~H., {et~al.} 2023, \mnras, 519, 2630, \dodoi{10.1093/mnras/stac3273}

\bibitem[{{Wylezalek} {et~al.}(2013){Wylezalek}, {Galametz}, {Stern}, {Vernet}, {De Breuck}, {Seymour}, {Brodwin}, {Eisenhardt}, {Gonzalez}, {Hatch}, {Jarvis}, {Rettura}, {Stanford}, \& {Stevens}}]{wylezalek13}
{Wylezalek}, D., {Galametz}, A., {Stern}, D., {et~al.} 2013, \apj, 769, 79, \dodoi{10.1088/0004-637X/769/1/79}

\bibitem[{{York} {et~al.}(2000){York}, {Adelman}, {Anderson}, {Anderson}, {Annis}, {Bahcall}, {Bakken}, {Barkhouser}, {Bastian}, {Berman}, {Boroski}, {Bracker}, {Briegel}, {Briggs}, {Brinkmann}, {Brunner}, {Burles}, {Carey}, {Carr}, {Castander}, {Chen}, {Colestock}, {Connolly}, {Crocker}, {Csabai}, {Czarapata}, {Davis}, {Doi}, {Dombeck}, {Eisenstein}, {Ellman}, {Elms}, {Evans}, {Fan}, {Federwitz}, {Fiscelli}, {Friedman}, {Frieman}, {Fukugita}, {Gillespie}, {Gunn}, {Gurbani}, {de Haas}, {Haldeman}, {Harris}, {Hayes}, {Heckman}, {Hennessy}, {Hindsley}, {Holm}, {Holmgren}, {Huang}, {Hull}, {Husby}, {Ichikawa}, {Ichikawa}, {Ivezi{\'c}}, {Kent}, {Kim}, {Kinney}, {Klaene}, {Kleinman}, {Kleinman}, {Knapp}, {Korienek}, {Kron}, {Kunszt}, {Lamb}, {Lee}, {Leger}, {Limmongkol}, {Lindenmeyer}, {Long}, {Loomis}, {Loveday}, {Lucinio}, {Lupton}, {MacKinnon}, {Mannery}, {Mantsch}, {Margon}, {McGehee}, {McKay}, {Meiksin}, {Merelli}, {Monet}, {Munn}, {Narayanan}, {Nash}, {Neilsen}, {Neswold}, {Newberg}, {Nichol}, {Nicinski},
  {Nonino}, {Okada}, {Okamura}, {Ostriker}, {Owen}, {Pauls}, {Peoples}, {Peterson}, {Petravick}, {Pier}, {Pope}, {Pordes}, {Prosapio}, {Rechenmacher}, {Quinn}, {Richards}, {Richmond}, {Rivetta}, {Rockosi}, {Ruthmansdorfer}, {Sandford}, {Schlegel}, {Schneider}, {Sekiguchi}, {Sergey}, {Shimasaku}, {Siegmund}, {Smee}, {Smith}, {Snedden}, {Stone}, {Stoughton}, {Strauss}, {Stubbs}, {SubbaRao}, {Szalay}, {Szapudi}, {Szokoly}, {Thakar}, {Tremonti}, {Tucker}, {Uomoto}, {Vanden Berk}, {Vogeley}, {Waddell}, {Wang}, {Watanabe}, {Weinberg}, {Yanny}, {Yasuda}, \& {SDSS Collaboration}}]{york00}
{York}, D.~G., {Adelman}, J., {Anderson}, John~E., J., {et~al.} 2000, \aj, 120, 1579, \dodoi{10.1086/301513}

\end{thebibliography}

\end{document}